# Tailors: New Music Timbre Visualizer to Entertain Music Through Imagery

## 음악의 음색을 강조한 시각화 시스템 개발: 심상 형성과 음악 향유 중심의 분석

# Contents

# Contents

# ABSTRACT

In this paper, I have implemented a timbre visualization system called 'Tailors.' Through the experiment with 27 MIR users, Tailors was found to be effective in conveying timbral warmth, brightness, depth, shallowness, hardness, roughness, and sharpness features of music compared to the only music condition and basic visualization. All scores of Tailors in the music imagery and music entertainment surveys were valued highest among the three conditions. Multiple linear regression analysis between timbre-imagery and imagery-entertainment shows significant and positive correlations. Coefficients comparing results from Fisher Transformation show that Tailors made users' music entertainment better through improved music visual imagery. The post-survey result represents that Tailors ranked first for the best timbre expression, music experience, and willingness to use it again. While some users felt a burden in the eye, Tailors left the future work of the data-driven approach of the mapping rule of timbre visualization to gain consent from many users. Furthermore, reducing timbre features to focus on features that Tailors can express well was also discussed, with future work of Tailors in a more artistic way using the sense of space.

# LIST OF TABLES

# LIST OF FIGURES

# I. INTRODUCTION

Music is a vital component in our lives, in the view of not only entertaining the music but also creating visual imagery and having a richer music experience. Through the music visual imagery, MIR (Music Information Retrieval) users can set the direction for their music entertainment. Because visual imagery in music can increase enjoyment from listeners [37], it is essential to form imagery in the best way. And it can be strengthened by the secondary creation of music, like music visualizations [33], because it has a strong connection between motor imagery in music visualization and the auditory imagery of the music [33, 35]. This gives importance to the philosophy of music visualization, to make MIR users entertained by the music well by the visualization. Music visualizations are created primarily based on what creators value the most.

Timbre, a complex and high-level music feature, has been neglected so far in music visualization. The recent trend of music visualization was focused on low-level features like the pitch and volume of the music. These features were simpler and relatively easier to express than other high-level features like timbre. The reason that timbre was hard to be represented by visuals was that there needed to be more on the criteria of the timbre. Many researchers have defined the timbral features of music into the kinds of instruments used in the music [26, 27, 28]. And this led to the exhaustion of ideas in music visualization because there needed to be more visual components to express the kinds of instruments. Semantic descriptors of timbre, on the other hand, were a breakthrough in the field of music visualization. Recently, more techniques are being used to express timbre using semantic descriptors [56], giving the advantage of being able to describe the subjective feeling of users, stretching out to the music visualization. However, compared to the importance of timbre, it is not widely used for music visualization and research on the good mapping rule using the timbral features [31]. It showed the possibilities and research needs about music visualization using timbre and the effects on the music visual imagery and entertainment.

In this paper, I have developed a web-based music visualization system using the timbral features of the music, called Tailors, to prove its effectiveness in timbre visualization compared to the condition of the only music and the basic visualization without timbral features. Furthermore, I want to identify the answer to these two research questions below by Tailors:

RQ1. Did timbral music visualization through Tailors well convey the timbral features of music?
RQ2. Did Tailors make music entertainment better through improved music visual imagery?

To prove these research questions, I conducted the main study of 27 participants listening to twenty pieces of music by three conditions (only music, basic visualization, and timbre visualization by Tailors). Each participant did the three kinds of the survey every time they experienced the music, producing 1,620 results for the total (27 participants * 20 pieces of music * 3 conditions). In addition, a demographic and a post-survey were included in the experiment to provide more insights into the results.

# II. RELATED WORKS & BACKGROUND

2. 1. Timbre

Timbre, also known as tone color, can show a unique spirit that music initially has. While we listen to music as a harmony of various components, timbre is a vital component that defines the music's overall mood and sensation. Different criteria defined timbre in multiple ways in the last decades [2]. Dolan [3] described timbre as a concept of discriminator of nonidentical sounds with similar pitch and loudness that are not the same. Wallmark [4] defined timbre as both a consequence of material sound and the inner sound of a listener, which is a subjective aesthetic judgment of minds. Siedenburg and McAdams [5] put together these opinions about timbre and presented timbre terms in four concepts. That timbre is a 1) perceptual attribute, 2) quality and a contributor to source identity, 3) functions on different scales of detail, and 4) property of fused auditory events. Because timbre depends very strongly on the acoustic properties of the music [1], identifying each content of acoustic properties is essential to knowing each music's timbral features. There were several attempts to set up the semantic descriptors for timbral features. Disley and Howard [7] gathered forty-five listeners' adjective words related to timbre. And they advanced their results by collecting words from musicians consistently used to describe the timbre of musical instruments in their following study [8]. Porcello [9] defined a taxonomy for timbre verbalizations from sound experts and distinguished these descriptors on vocal and non-vocal timbres. Because timbral features are multidimensional [19, 22], it is essential to clarify each dimension and its meaning to understand timbre. Pearce et al. [6] collected and grouped 145 timbral attributes considering group discussion results and filtering by search term frequencies. Bismarck [11] stated that sharpness is a timbral component that distinguishes the pitch and loudness of verbal attributes. Stark [12] mentioned timbral brightness in singing voice pedagogy, highlighting it in vocal resonances, a feature related to high frequencies' spectral energy [21]. Pearce et al. [14] have built a perceptual model for hardness because it was undervalued in its importance in acoustic musical acoustics. Pressnitzer et al. [16] claimed that roughness plays an important role in musical tension perception because it is proposed on a sensory basis attribute. Vassilakis [17] also argued the importance of auditory roughness in the aesthetic aspect of music. Wen [20] defined depth as a type of music with a poetic and dreamy mood and found that it relates to different music features like tempo, MFCC, etc. For the warmth of the music, it was found that it is highly associated with pitches, which is a component that can give a pleasant sound sensation [21]. Due to this multidimensionality of timbral features, the need to classify these features automatically using the feature extraction model has arisen. Although a lot of research organized multiple timbres of instruments in a piece of single music [26, 27, 28], there was little timbral discrimination on a single musical instrument. Loureiro et al. [24] investigated classification methods for a single sound using a clustering algorithm but with no labels of timbre descriptors. Recently, Oliván and Blázquez [23] developed a multi-head attention-based model that classifies different timbres of the instruments. Although this model achieved an overall F1 value of 0.62, the model was not for organizing vocal timbres. Sha and Yang [25] build an automatic classifier on singing voice using 387 popular Chinese songs, achieving 79.84% accuracy; however, with no instrument timbres.

2. 2. Music Visualization and Music Visual Imagery

To express mood, emotion, and other audio features, music visualizations were used to convey these sentiments with information to MIR users. Music visualizations are direct reflections of musical representations with good explanatory power [30], which can deepen our understanding of musical pieces [29]. For those who are not specialists in music, visualization can help them comprehend the music's feelings and expressions with various effects [32]. In addition, music visualization can provide an intuitive visual presentation of music to MIR users, which is highly related to synesthesia [33]. Synesthesia, which can be generated by music, heavily relies on cross-modal association while music listening experience [34]. It happens because there's a strong connection between auditory and motor imagery, showing similar consequences in musical performance and imagery function neurologically [35, 41]. It shows that music visualization can cause synesthesia in the interactive visual and audio process [33]. Even though visual imagery could be generated with most of the expressive features of music [36, 37, 38], timbral features are an essential dimension of developing music visual imagery rather than other musical attributes [40]. Halpern et al. [41] found that perception and visual imagery access a similar cognitive representation of timbre and that timbre imagery activated other auditory areas connected to the visual imagery. Bailes [39] experimented with vivid timbre perception and imagery for music. She found that the participants could internalize timbral features of music using the ability to discriminate timbres through the study. This shows that timbre is a crucial characteristic of the sound that can generate visual music imagery.

Consequently, users' music visual imagery, especially the one that is evoked by timbre, can be strengthened by music visualizations [36], leading them to enhance their musical experience with enjoyment [37]. According to a recent music visualization survey paper [31], it was found that timbre was timidly used in features for music visualizers despite the importance of music visual imagery and enjoyment. Smith and Williams [42] defined timbre as the most complex musical attribute to visualize because it depends on many factors. However, there were several attempts to imagine music with timbral features. Li et al. [43] employed three pairs of images containing the piano timbre's brightness, shape, and size. They have defined soft timbres as round shapes with cold colors and harsh timbres as angular shapes with warm colors. Giannakis [44] mapped timbral features such as sharpness, compactness, and sensory dissonance to the visual textures and developed a music visualization system called Sound Mosaics. Although Sound Mosaics was comprehensible in the timbral aspect to users, it had limitations on the small sample size and low statistical significance of the experiment result. Siedenburg [45] also explored various representations of real-time timbral visualization using the music programming environment, leaving the future work of music visualization needing some creativity.

# III. STUDY

3. 1. Identity of Tailors

The identity of Tailors came from semantic descriptors of the timbral features of the music and applying it to the music visualizations. Besides the discrimination on the instrument, these descriptors could well convey the specific meaning and mood of the timbral features. As mentioned in the Related Work and Background Section, there are several timbral visualizations in the world [44, 45], but it has yet to strengthen the music visual imagery of the user. If Tailors could support the music visual imagery of the users, those users could more entertain by the music by using the imagery that pops up in their minds. I wanted to determine this process as 'tailoring' of the musical piece by timbre visualization to each user, making the goods fit a given target. Through semantic descriptors as a mapping component for timbre visualization, Tailors were able to help users invent music-visual imagery and lead it to music entertainment.

To express and reflect the timbral features in the music visualization well, I conceived the art trend of Impressionism to the Tailors's visualization part. Impressionism is an art style that nicely captures the instantaneous changes in the shape of natural objects according to the transformation of the light [46]. Impressionist artists used color segmentation techniques to give a sense of change in form, which shows multiple colors to express one object. I thought these impressionist-style visualizations could well represent the timbral features of the music through color, tone, and texture. In addition, impressed by the nature and the sound artwork by Anna Marinenko [47], Tailors took the view that music visualization should express the whole flow of the sound by nature. By this, I used three types of nature: cloud, water, and ice in the Tailors. Additionally, for easy mapping and to make more difference in the numbers from the timbral model, the mapping rule in Tailors used min-max normalization. Appendix A shows the user interface of Tailors.

3. 2. System Design and Mapping Rule of Tailors

3. 2. 1. System Design

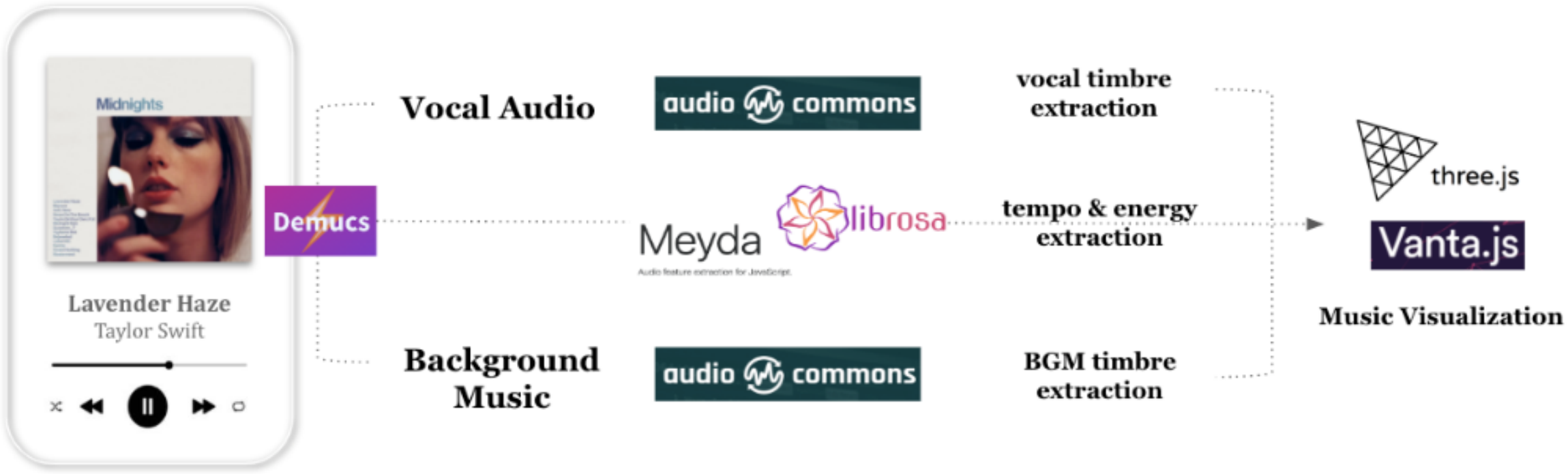

Figure 1. Overall System Structure of Tailors

Figure 1 shows the whole system structure for the Tailors. First, to deliver the timbral feature through the visualization well, I separated the original music into the vocal audio and the background music using the audio source separation library called demucs [48]. Then, because visualization elements that can better express vocal and background music are different, I separated the rule and expressed timbre components on the Tailors. Then, due to the difference in visualization components and methods and the applied timbral features in the vocal and the background, I macroscopically divided it into the object (vocal audio) and the background, which shows the flow (background music). Then, using the timbral model from audio commons [49], each timbral feature for both vocal audio and the background music expresses the original sound well was extracted. For the vocal audio, timbral features of roughness [13], sharpness [11], and warmth [52] were extracted. And for the background music, timbral features of roughness [16, 17], depth [50], brightness [15], hardness [10], and warmth [51] were extracted. For the visualization of the timbral features, three.js [53], a javascript library, was used for the object - the vocal visualization. And a Vanta.js [54] for the background visualization to represent 3D animated backgrounds.

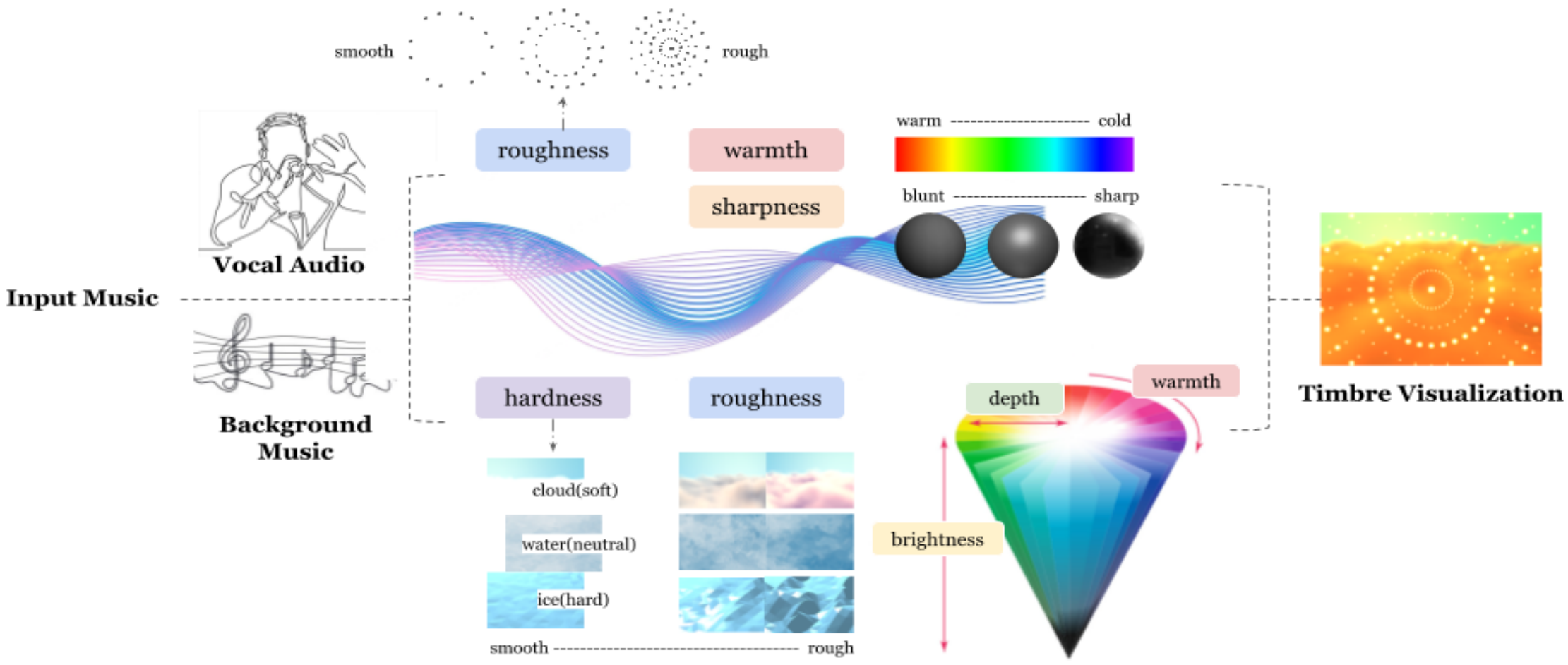

Figure 2. Timbre Visualization Mapping Rule of Tailors

3. 2. 2. Mapping Rule for Vocal Timbre

As shown in Figure 2, visualization components and the timbral mapping rule in Tailors are as follows. The object, representing the vocal audio, is a combination of small sphere objects forming a giant sphere. These small sphere objects were controlled to express the roughness of the vocal sound, representing rough texture when it comes together and creates a big sphere object. By changing the view of the user's sight in the visualization, Tailors intended users to feel either rough or smooth. The sharpness and the warmth of the vocal timbre express the object's texture and the hue of the color, respectively. If the sharpness of the vocal is high, the texture of the object gets closer to the metal texture and gets closer to the plain texture in vice versa. Through the warmth score of the vocal, the object's color gets close to the warm color (*e.g.*, red, orange, or yellow) when it gets a high score in warmth and gets close to the cold color (*e.g.*, green, blue, or violet) in the vice versa.

3. 2. 3. Mapping Rule for Background Timbre

The background is divided into three categories based on the score for the timbral hardness of the background music. Because timbral hardness represents the strongness of the music, I mapped the strong timbre group into the group of ice, the neutral group to the water, and the soft timbre group to the cloud. Based on these criteria, each natural object gets to express the roughness of the background music. In addition, the same rule as the vocal object applies to the background color- timbral warmth to the background color's hue, and two additional rules for the background- timbral brightness to the background color's value and the timbral depth to the its saturation.

## 3. 3. Method

### 3. 3. 1. Participants and Procedure

| Category | | N (Total 27) | Percentage (%) |
|---|---|---|---|
| **Gender** | | | |
| | Female | 16 | 59.30% |
| | Male | 11 | 40.70% |
| **Age** | | | |
| | 18-23 | 4 | 17.90% |
| | 24-29 | 19 | 64.20% |
| | 30-35 | 4 | 17.90% |
| **Interest in Music Listening** | | | |
| | Very Interested | 11 | 40.70% |
| | Somewhat Interested | 12 | 44.40% |
| | Neutral | 3 | 11.10% |
| | Not Very Interested | 0 | 0% |
| | Not at all Interested | 1 | 3.70% |
| **Frequency of Music Listening** | | | |
| | More than 5 hours per day | 2 | 7.40% |
| | 3 to 5 hours per day | 3 | 11.10% |
| | 1 to 2 hours per day | 17 | 63.00% |
| | Less than 1 hour per day | 5 | 18.50% |
| **Favorite Genre of Music (Duplicate answer possible)** | | | |
| | Classic | 6 | 22.20% |
| | POP | 16 | 59.30% |
| | CCM | 3 | 11.10% |
| | Jazz | 2 | 7.40% |
| | K-POP | 17 | 63.00% |
| | OST (Movie, Drama) | 12 | 44.40% |
| | Ballad | 19 | 70.40% |
| | R&B | 7 | 25.90% |
| | Hip-hop | 12 | 44.40% |
| | Trot | 1 | 3.70% |
| | Indie | 8 | 29.60% |
| | EDM | 2 | 7.40% |
| | Country | 1 | 3.70% |
| | Folk | 1 | 3.70% |
| | Korean Traditional Music | 1 | 3.70% |
| | Alternative Rock & Band | 1 | 3.70% |
| | Hymn | 1 | 3.70% |
| **Interest in Music Visualization** | | | |
| | Very Interested | 1 | 3.70% |
| | Somewhat Interested | 1 | 3.70% |
| | Neutral | 0 | 0.00% |
| | Not Very Interested | 3 | 11.10% |
| | Not at all Interested | 22 | 81.50% |

Table 1. Demographic information of the participants

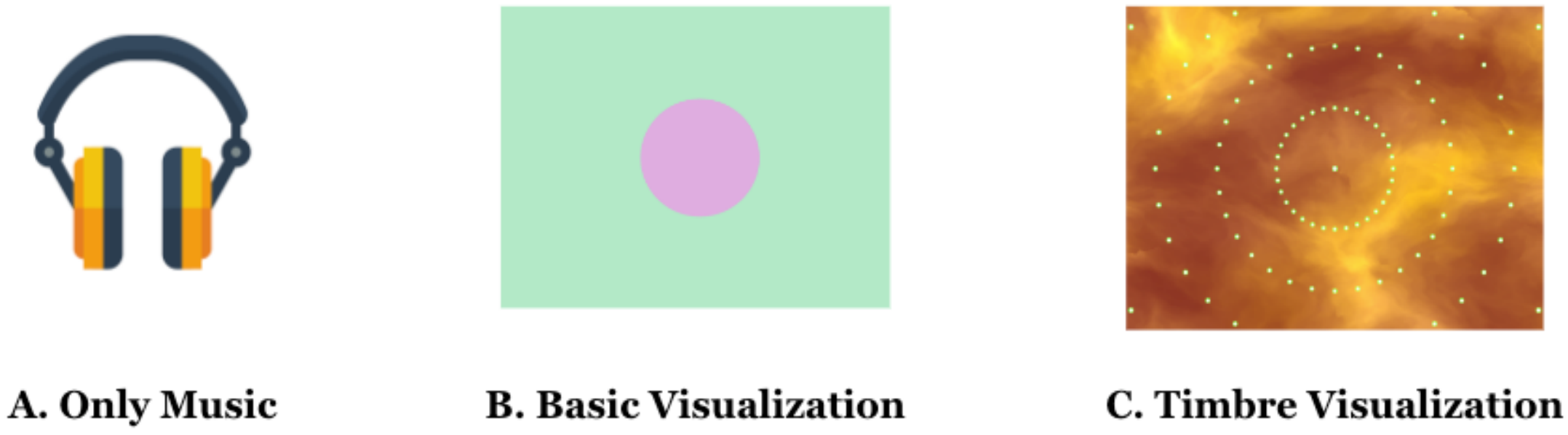

Figure 3. Three music conditions (A, B, C) in experiment

27 participants who enjoyed listening to music daily were recruited for the experiment. I gathered the demographic questionnaire of the users before the experiment. Table 1 shows the result of the demographics of the participants. For the procedure of the main study, every participant got to experience twenty pieces of POP music under three conditions (Figure 3) and do surveys after listening to a piece of music in one condition. These three conditions represents **A. Only Music**, **B. Basic Visualization**, and **C. Timbre Visualization (Tailors)**. In the main study they were counter-balanced by the Latin-Square Design, additionally shuffling the twenty pieces of music on every user. Detailed information about the twenty pieces of music used in the experiment is in the Appendix B. Furthermore, because the experiment procedure went on a web-based, every participant was given the website link to get access, and they were paid each ₩30,000 for participating in the experiment. The overall experiment procedure went approximately two hours per person. See Appendix C for the visualized output of the Tailors.

3. 3. 2. Materials and Metrics

| | **Questionnaire in the Timbre Survey** | **Timbre Feature** |
|---|---|---|
| **1** | I felt the power of the timbre hard. | hard |
| **2** | I felt the power of the timbre soft. | soft |
| **3** | I felt the timbre complicated and deep. | deep |
| **4** | I felt the timbre simple and shallow. | shallow |
| **5** | I felt the timbral brightness. | bright |
| **6** | I felt the timbral darkness. | dark |
| **7** | I felt the timbral warmness. | warm |
| **8** | I felt the timbral coldness. | cold |
| **9** | I felt the timbral roughness. | rough |
| **10** | I felt the timbral smoothness. | smooth |
| **11** | I felt the timbral sharpness. | sharp |
| **12** | I felt the timbral bluntness. | blunt |

Table 2a. Music Timbre Survey Questionnaires

| | Questionnaire in the Imagery Survey | Imagery Feature |
|---|---|---|
| 1 | I felt harmonious and balanced. | flow |
| 2 | I felt the power that music gives me. | force |
| 3 | I was able to become one with the music. | interior |
| 4 | I was able to move my body to the rhythm. | movement |
| 5 | I wanted to wander and travel around. | wandering |

Table 2b. Music Imagery Survey Questionnaires

| | Questionnaire in the Entertainment Survey | Entertainment Feature |
|---|---|---|
| 1 | I felt a shiver run through my body. | stimulated |
| 2 | I wanted to dance around. | dancing |
| 3 | I was able to feel entertained. | entertained |
| 4 | I was able to feel energized. | energized |
| 5 | I was moved. | moving |
| 6 | I was able to feel animated. | animated |
| 7 | I got excited. | excited |
| 8 | I was able to feel the rhythm well. | rhythm |

Table 2c. Music Entertainment Survey Questionnaires

Due to the research questions being focused on figuring out the Tailors' effect on delivering timbral features, and creation of imagery and entertainment of the music through it, the main study compared timbral music visualization with the condition of the music only and the basic visualization with volume(energy) of the music but without timbral features. I have defined the three surveys for music timbre, imagery, and entertainment to get from the users, and Table 2 shows the questionnaire of each survey; Table 2a represents the twelve 7-point Likert scale questions for the Music Timbre Survey (hard, soft, deep, shallow, bright, dark, warm, cold, rough, smooth, sharp, and blunt). Table 2b shows the music imagery survey's five 7-point Likert scale questions (flow, force, interior, movement, and wandering) from GEMMES [55], which is a questionnaire for the music metaphor. And Table 2c shows the eight 7-point Likert scale questions (stimulated, dancing, entertained, energized, moving, animated, excited, and rhythm) for the Music Entertainment Questionnaire [55].

# IV. RESULTS

## 4. 1. Did timbral music visualization through Tailors well convey the timbral features of music? (RQ1)

### 4. 1. 1. Overall Results of the Timbre Survey

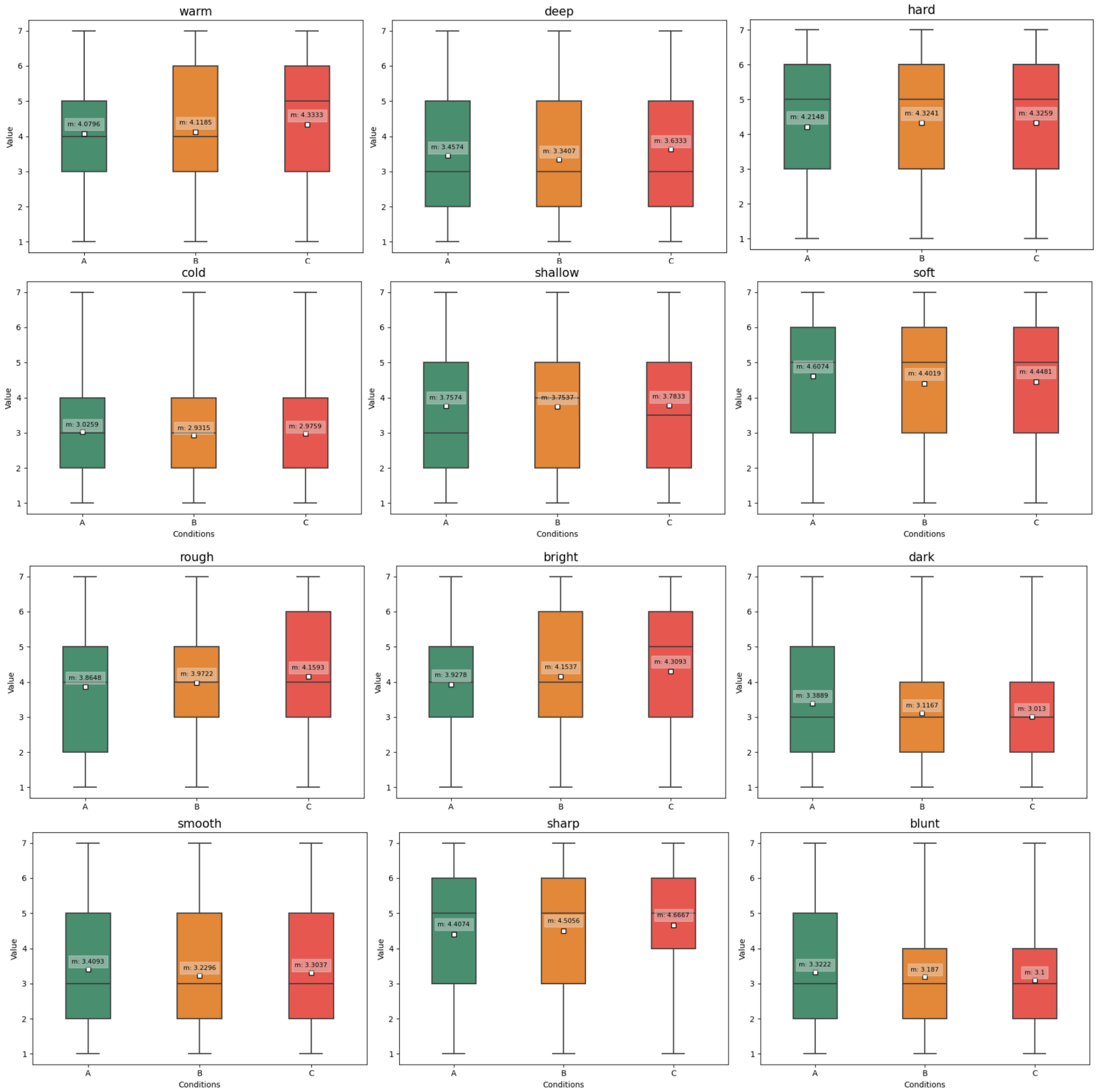

Figure 4. Boxplot for each timbral feature value after experiencing Tailors

To figure out that timbre visualization through Tailors was effective for conveying timbral features of the music well, I compared three conditions of the music: **A. Only Music**, **B. Basic Visualization**, and **C. Timbre Visualization (Tailors)**. First, I conducted a Kruskal-Wallis test to compare three conditions, but no significance was found. However, as shown in Figure 4, it was found that seven features (warm, bright, deep, shallow, hard, rough, and sharp) among twelve timbral features were most well delivered by timbre visualization among the three conditions. On the other hand, for three features (cold, smooth, soft), it was found that it conveyed its timbral features better than basic visualization but less than only music. And for the last two elements (dark, blunt), timbre visualization didn't deliver the best of it.

Appendix D1 shows that there were combinations that had significant differences in the Wilcoxon sign-rank test between A. only music and C. timbre visualization and A. only music and B. basic visualization. In the comparison of only music versus timbre visualization, I figured out that timbre visualization better conveyed significantly bright, warm, and rough features of music than only music condition. Although other timbral features had no significance in difference, hard, deep, shallow, and sharp timbral features better delivered timbral features in timbre visualization than only music. However, dark, blunt features went vice versa. In the comparison of only music versus basic visualization, timbral elements of dark and soft were better conveyed in only music than in basic visualization. And other parts were shown as insignificant, while deep, cold, smooth, and blunt elements went better in only music. While hard, shallow, bright, warm, rough, and sharp went better in basic visualization.

#### 4. 1. 2. Timbre Survey Results By Groups

To learn more specifically about the overall results, I divided five groups using the original music information and the results of the timbre survey of 27 users. I wanted to figure out what timbral features Tailors delivered well or not, so I compared the timbre survey to each piece of music, with the comparison of music's timbral feature of its own and the representation in Tailors. The first group represents examples of good representation and good results of Tailors, which shows well-made visualization by timbral features of music and the cases where users could be well-delivered timbre. The second group is the example of lousy representation and destructive results of Tailors, which is vice versa to the first group. The third group represents neutral examples, which it well delivered the music's timbre but the result was neither good nor bad. And the fourth group shows the counter-example of Tailors, in which timbre visualization has the lowest score in the feature of the opposite of the timbre that the music initially has. Finally, the last group shows the extraordinary example, which shows the reversed output of the expected results- which shows the user survey was dependent on the expression of the timbral visualization, little to do with the original timbre that music has. I counted the number of instances for each group in each timbral feature and found the insights below.

##### Group of Good Results

Of the total 240 cases (20 pieces of music * 12 timbral features), I found 79 cases, which is 32.91% of the total found to be a good result. The good result, representing timbral features of the music is well delivered in the Tailors, took up a large part in these features: deep (14), warm (12), rough (11), bright (9), soft (7), sharp (7). Comparing this to the overall results of the timbre survey, all the features in the group of good results except timbral softness and shallowness were the features that were the most well delivered by timbre visualization among the three conditions. This cross-validates the overall results of the timbre survey, which shows that Tailors effectively delivered deep, warm, rough, bright, and sharp features of the timbre in music.

##### Group of Bad Results

On the other hand, 31 out of the 240 cases, 12.91% of the total, were a bad result, representing the timbral features of the music not well delivered in the Tailors. The number of bad results in timbral darkness (5) and bluntness (6) was the highest among the twelve features. Comparing this to the overall results of the timbre survey, all the features in the group of bad outcomes were the features that were the most poorly delivered by timbre visualization among the three conditions. This also cross-validates the overall results of the timbre survey, which shows that Tailors was weak in providing dark and blunt features of the timbre in music.

Group of Neutral Results
65 out of 240 cases, 27.08% of the total, were classified into neutral results, which was neither good nor bad. I found timbral hardness (12), shallowness (9), sharpness (8), and smoothness (7) had a large portion in the neutral result.

Group of Counter Results
38 out of 240 cases, which is 15.83% of the total, were classified into the counter results, in which Tailors has the lowest score in the feature because it was an opposite feature of the timbre that the music initially had. I found timbral softness (7), darkness (6), coldness (5), and smoothness (5) had the most portion in the counter results. And also found that timbral coldness and smoothness were the features that had a lower score than the only music condition but higher than the basic visualization. And because the timbral darkness in Tailors was the weakest among the three conditions, the low score of timbral darkness is that there was more bright music than dark music among the 20 pieces of music.

Group of Extraordinary Results
27 out of 240 cases, 11.25% of the total, were classified into extraordinary results, which was an unexpected result compared to the original timbral feature of the music. I found timbral shallowness (5), coldness (4), sharpness (4), and roughness (4) had the most portion in the extraordinary results. And I also figured out that for timbral shallowness, the number of cases with good results was only one. This shows that although the timbral shallowness score in Tailors was highest among the three conditions, it may need to be more balanced by the extraordinary result, not by the good result. This represents that users got to feel timbral shallowness more than the timbre of the actual music through Tailors.

4. 2. Did Tailors make music entertainment better through improved music visual imagery? (RQ2)

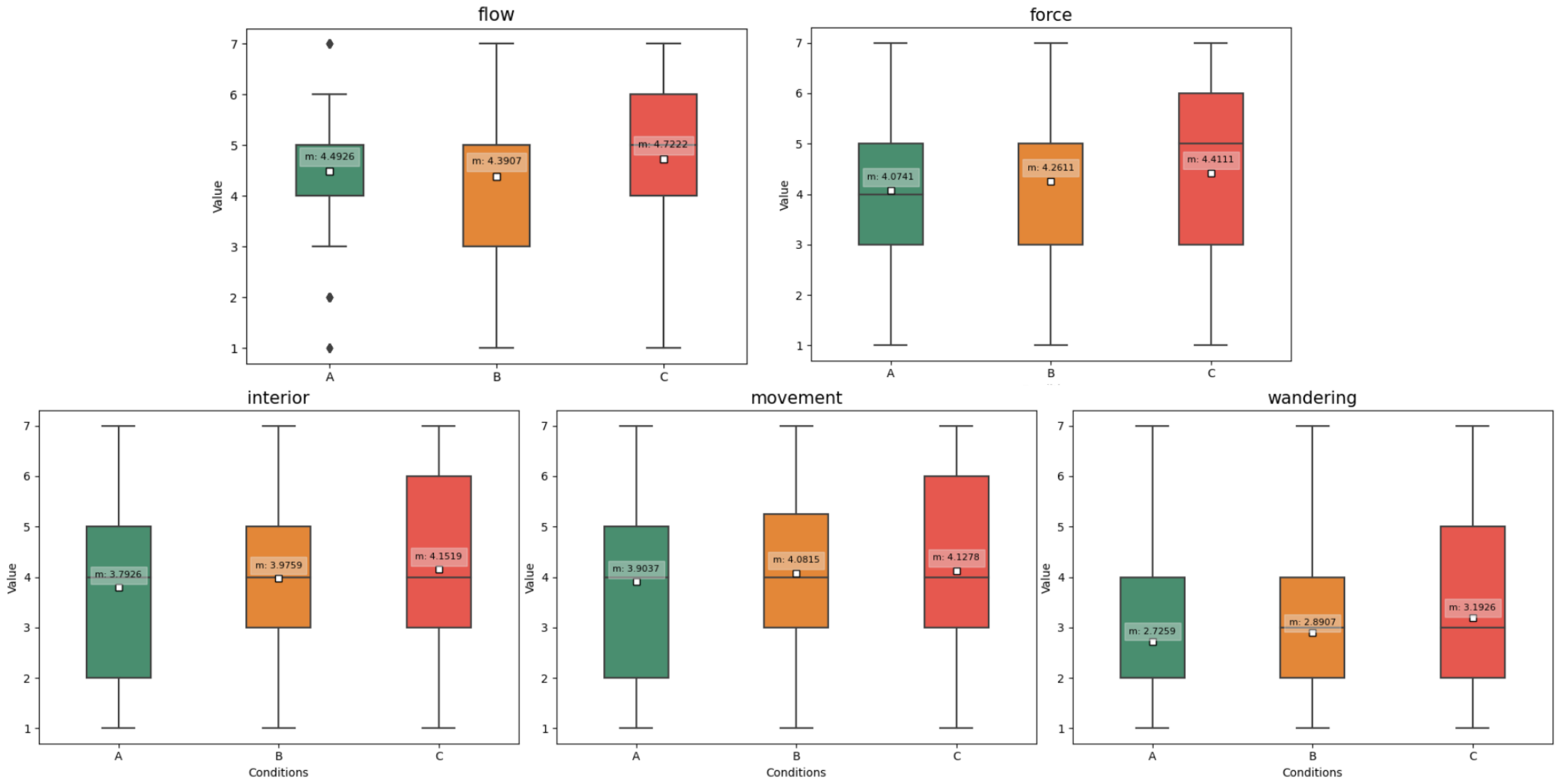

Figure 5. Boxplot for each imagery feature value after experiencing Tailors

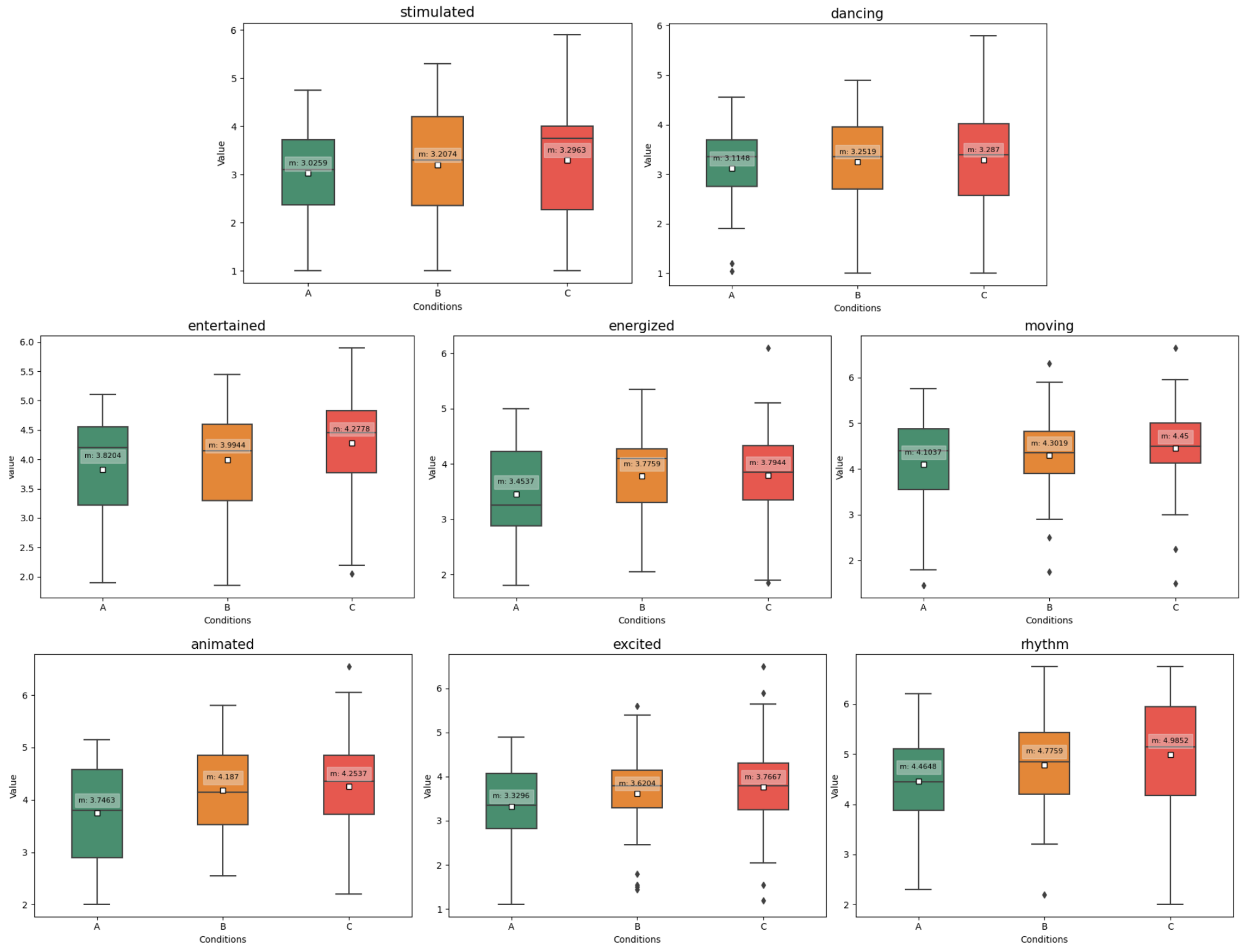

Figure 6. Boxplot for each entertainment feature value after experiencing Tailors

4. 2. 1. Overall Results of the Imagery Survey

Figure 5 shows the comparison results of the music visual imagery survey. Among the three conditions, timbre visualization was the most effective in all imagery aspects (flow, force, interior, movement, and wandering). Because it didn't show any significance in the Kruskal-Wallis test, I conducted a Wilcoxon sign-rank test between three combinations. Appendix D2 shows that varieties significantly differed in the Wilcoxon sign-rank test. Although all imagery aspects were highest in Tailors, there were significant differences in flow (mean score: 4.7222) and wandering (mean score: 3.1926) in timbre visualization compared to the basic visualization (flow mean score: 4.3907, wandering mean score: 2.8907). And it also turned out that the wandering aspect was significantly higher in timbre visualization (mean score: 3.1926) than the only music (mean score: 2.7259). In comparing basic visualization and only music, every aspect of imagery turned out there's no significance while showing an element of flow turned out to be higher than the basic visualization.

4. 2. 2. Overall Results of the Entertainment Survey

Figure 6 shows the comparison results of the music entertainment survey. Among the three conditions, timbre visualization is the most effective in all entertainment aspects (stimulated, dancing, entertained, energized, moving, animated, excited, and rhythm). Because it didn't show any significance in the Kruskal-Wallis test, I conducted a Wilcoxon sign-rank test between three combinations. Appendix D3 shows that varieties significantly differed in the Wilcoxon sign-rank test. Although all entertainment aspects were highest in Tailors, there were significant differences in entertain aspect (mean score: 4.2778) in timbre visualization compared to the basic visualization (mean score: 3.9944). Compared with the only music condition, there was a significant difference in entertained (Tailors mean score: 4.2778, only music mean score:3.8204), animated (Tailors mean score: 4.2537, only music mean score:3.7463), excited (Tailors mean score: 3.7667, only music mean score:3.3296), and rhythm (Tailors mean score: 4.9852, only music mean score:4.4648). Compared to the only music condition and basic visualization, basic visualization was significantly higher than the only music condition in energized, animated, and exciting aspects.

4. 2. 3. Multiple Linear Regression Analysis on Tailors

To determine which components of timbre features affected which imagery and entertainment aspects of music, I conducted multiple linear regression analysis using OLS(Ordinary Least Squares) between twelve timbre features and five elements of the imagery, and eight elements of music entertainment. The range of the coefficient is from -1 to +1, which shows strong positive correlation between variables closer to +1, and vice versa to −1. To determine effective timbral features in each music imagery and entertainment feature, I compared coefficients in multiple regression equations, and preprocessed each survey points to remove the multicollinearity between the variables.

| IV (Timbre) | DV (Imagery) | coefficients | std_err | t_value | p_value | R-squared | Adj. R-squared | F-Statistic |
|---|---|---|---|---|---|---|---|---|
| hard | flow | 0.2738 | 0.2437 | 1.1235 | 0.2801 | 0.615 | 0.286 | 1.866 |
| soft | flow | -0.0493 | 0.2953 | -0.1669 | 0.8698 | 0.615 | 0.286 | 1.866 |
| deep | flow | -0.1238 | 0.3272 | -0.3784 | 0.7108 | 0.615 | 0.286 | 1.866 |
| shallow | flow | 0.1457 | 0.4082 | 0.3569 | 0.7265 | 0.615 | 0.286 | 1.866 |
| bright | flow | -0.0587 | 0.333 | -0.1762 | 0.8627 | 0.615 | 0.286 | 1.866 |
| dark | flow | -0.6351 | 0.3292 | -1.9292 | 0.0742 | 0.615 | 0.286 | 1.866 |
| warm | flow | 0.3726 | 0.347 | 1.074 | 0.301 | 0.615 | 0.286 | 1.866 |
| cold | flow | 0.3282 | 0.3205 | 1.024 | 0.3232 | 0.615 | 0.286 | 1.866 |
| rough | flow | 0.2802 | 0.3716 | 0.754 | 0.4633 | 0.615 | 0.286 | 1.866 |
| smooth | flow | 0.169 | 0.3523 | 0.4797 | 0.6389 | 0.615 | 0.286 | 1.866 |
| sharp | flow | 0.2765 | 0.5302 | 0.5215 | 0.6102 | 0.615 | 0.286 | 1.866 |
| blunt | flow | 0.1497 | 0.4932 | 0.3034 | 0.766 | 0.615 | 0.286 | 1.866 |
| hard | force | 0.4779 | 0.1691 | 2.8255 | 0.0135 | 0.815 | 0.656 | 5.131 |
| soft | force | 0.0455 | 0.2049 | 0.2219 | 0.8276 | 0.815 | 0.656 | 5.131 |
| deep | force | -0.5263 | 0.227 | -2.3181 | 0.0361 | 0.815 | 0.656 | 5.131 |
| shallow | force | -0.0252 | 0.2833 | -0.089 | 0.9304 | 0.815 | 0.656 | 5.131 |
| bright | force | 0.2806 | 0.2311 | 1.2145 | 0.2446 | 0.815 | 0.656 | 5.131 |
| dark | force | -0.2563 | 0.2284 | -1.1219 | 0.2808 | 0.815 | 0.656 | 5.131 |
| warm | force | 0.1611 | 0.2408 | 0.669 | 0.5144 | 0.815 | 0.656 | 5.131 |
| cold | force | 0.4237 | 0.2224 | 1.9046 | 0.0776 | 0.815 | 0.656 | 5.131 |
| rough | force | 0.4832 | 0.2579 | 1.8738 | 0.082 | 0.815 | 0.656 | 5.131 |
| smooth | force | 0.2476 | 0.2444 | 1.0129 | 0.3283 | 0.815 | 0.656 | 5.131 |
| sharp | force | 0.0257 | 0.368 | 0.0698 | 0.9453 | 0.815 | 0.656 | 5.131 |
| blunt | force | -0.0441 | 0.3423 | -0.129 | 0.8992 | 0.815 | 0.656 | 5.131 |
| hard | interior | 0.5569 | 0.1784 | 3.1212 | 0.0075 | 0.794 | 0.617 | 4.494 |
| soft | interior | 0.2463 | 0.2162 | 1.1396 | 0.2736 | 0.794 | 0.617 | 4.494 |
| deep | interior | -0.4004 | 0.2395 | -1.6719 | 0.1167 | 0.794 | 0.617 | 4.494 |
| shallow | interior | -0.3034 | 0.2988 | -1.0153 | 0.3272 | 0.794 | 0.617 | 4.494 |
| bright | interior | 0.1443 | 0.2437 | 0.592 | 0.5633 | 0.794 | 0.617 | 4.494 |
| dark | interior | -0.292 | 0.241 | -1.2118 | 0.2456 | 0.794 | 0.617 | 4.494 |
| warm | interior | 0.0647 | 0.254 | 0.2549 | 0.8025 | 0.794 | 0.617 | 4.494 |
| cold | interior | 0.7377 | 0.2346 | 3.1442 | 0.0072 | 0.794 | 0.617 | 4.494 |
| rough | interior | 0.667 | 0.272 | 2.4523 | 0.0279 | 0.794 | 0.617 | 4.494 |
| smooth | interior | 0.486 | 0.2578 | 1.8847 | 0.0804 | 0.794 | 0.617 | 4.494 |
| sharp | interior | -0.2274 | 0.3881 | -0.586 | 0.5672 | 0.794 | 0.617 | 4.494 |
| blunt | interior | -0.5799 | 0.361 | -1.6063 | 0.1305 | 0.794 | 0.617 | 4.494 |
| hard | movement | 0.6692 | 0.1616 | 4.1415 | 0.001 | 0.831 | 0.686 | 5.733 |
| soft | movement | 0.2058 | 0.1958 | 1.0514 | 0.3109 | 0.831 | 0.686 | 5.733 |
| deep | movement | -0.119 | 0.2169 | -0.5487 | 0.5919 | 0.831 | 0.686 | 5.733 |
| shallow | movement | 0.2842 | 0.2707 | 1.05 | 0.3115 | 0.831 | 0.686 | 5.733 |
| bright | movement | 0.0078 | 0.2208 | 0.0353 | 0.9724 | 0.831 | 0.686 | 5.733 |
| dark | movement | -0.4926 | 0.2182 | -2.2572 | 0.0405 | 0.831 | 0.686 | 5.733 |
| warm | movement | -0.0814 | 0.23 | -0.3537 | 0.7288 | 0.831 | 0.686 | 5.733 |
| cold | movement | 0.5876 | 0.2125 | 2.765 | 0.0152 | 0.831 | 0.686 | 5.733 |
| rough | movement | 0.6651 | 0.2464 | 2.6997 | 0.0173 | 0.831 | 0.686 | 5.733 |
| smooth | movement | 0.1293 | 0.2335 | 0.5536 | 0.5886 | 0.831 | 0.686 | 5.733 |
| sharp | movement | -0.4375 | 0.3515 | -1.2445 | 0.2338 | 0.831 | 0.686 | 5.733 |
| blunt | movement | -0.4756 | 0.327 | -1.4545 | 0.1679 | 0.831 | 0.686 | 5.733 |
| hard | wandering | 0.1795 | 0.2154 | 0.8333 | 0.4187 | 0.7 | 0.442 | 2.717 |
| soft | wandering | 0.132 | 0.261 | 0.5058 | 0.6209 | 0.7 | 0.442 | 2.717 |
| deep | wandering | 0.0479 | 0.2891 | 0.1657 | 0.8708 | 0.7 | 0.442 | 2.717 |
| shallow | wandering | -0.2254 | 0.3608 | -0.6248 | 0.5422 | 0.7 | 0.442 | 2.717 |
| bright | wandering | -0.1087 | 0.2942 | -0.3694 | 0.7173 | 0.7 | 0.442 | 2.717 |
| dark | wandering | -0.3357 | 0.2909 | -1.1541 | 0.2678 | 0.7 | 0.442 | 2.717 |
| warm | wandering | 0.0725 | 0.3066 | 0.2364 | 0.8165 | 0.7 | 0.442 | 2.717 |
| cold | wandering | 0.3854 | 0.2833 | 1.3605 | 0.1952 | 0.7 | 0.442 | 2.717 |
| rough | wandering | 0.7479 | 0.3284 | 2.2778 | 0.039 | 0.7 | 0.442 | 2.717 |
| smooth | wandering | 0.4513 | 0.3113 | 1.4498 | 0.1691 | 0.7 | 0.442 | 2.717 |
| sharp | wandering | 0.0948 | 0.4686 | 0.2023 | 0.8426 | 0.7 | 0.442 | 2.717 |
| blunt | wandering | -0.1235 | 0.4358 | -0.2835 | 0.781 | 0.7 | 0.442 | 2.717 |

Table 3a. Multiple Linear Regression Result of Timbre and Imagery from Tailors

Table 3a shows the multiple linear regression analysis results between twelve timbre features and five imagery aspects in timbre visualization. Red letters in each table represent a significant positive predictor variable with a predicted value. Blue letters in each table represent a significant negative predictor variable with a predicted value. Except for the flow aspect, which had no significant timbral feature (F(12, 14)=1.866, p>0.05), I was able to observe significant effects for the last four imagery features. For the force aspect, the timbre feature of hard turned out to be a significant and positive main cause (F(12, 14)=5.131, p=0.0135). The regression coefficient of hard was 0.4779, which was found to be a slightly weak significant positive predictor of force (t(14)=2.8255, p=0.0135). On the other hand, the deep timbral feature turned out to be a negative predictor (F(12, 14)=5.131, p=0.0361), showing its coefficient of -0.5263 (t(14)=-2.3181, p=0.0361). In the interior aspect, the hard, cold, and rough timbre features turned out to be significant and positive causes for the effect (F(12, 14)=4.4940, p<0.05). The coefficient of cold was 0.7377 (t(14)=3.1442, p=0.0072), and rough was 0.6670 (t(14)=2.4523, p=0.0279), which shows cold and rough feature has strong positive correlations with the interior. Also, the coefficient of hard was 0.5569, which also shows medium positive correlations with the interior (t(14)=3.1212, p=0.0279). The timbre feature of hard, cold, and rough had a significant positive effect on the movement aspect, too (F(12, 14)=5.7330, p<0.05). The coefficient of hard timbre was 0.6692 (t(14)=4.1415, p=0.0010), rough was 0.6651 (t(14)=2.6997, p=0.0173), and cold was 0.5876 (t(14)=2.6997, p=0.0173). This result shows that hard, rough, and cold timbral features were strong and significant positive predictors for movement imagery. Lastly, for the wandering aspect, the timbre feature of rough was found to be a strong, significant, and positive factor (F(12, 14)=2.7170, p<0.05). The coefficient of the rough feature of timbre was 0.7479, which shows a strong positive correlation between the imagery aspect of wandering. To summarize, except for the flow aspect of music visual imagery, users of timbre visualization by Tailors were positively affected by timbral features of hard, cold, and rough, which help them internalize with music (interior) and were able to move their bodies by the rhythm (movement). In feeling the force of the music (force) given to the user, a timbral feature of hard was a slightly weak positive factor. Conversely, the timbral feature of deep turned out to be a negative factor for the imagery aspect of the force. For the feeling of wandering and traveling around the world (wandering), the rough feature of the timbre was a strong positive cause.

| IV (Imagery) | DV (Entertainment) | coefficients | std_err | t_value | p_value | R-squared | Adj. R-squared | F-Statistic |
|---|---|---|---|---|---|---|---|---|
| flow | stimulated | -0.0453 | 0.2119 | -0.2139 | 0.8327 | 0.707 | 0.637 | 10.11 |
| force | stimulated | -0.746 | 0.2903 | -2.57 | 0.0178 | 0.707 | 0.637 | 10.11 |
| interior | stimulated | 0.8757 | 0.2918 | 3.0006 | 0.0068 | 0.707 | 0.637 | 10.11 |
| movement | stimulated | 0.4512 | 0.322 | 1.4013 | 0.1757 | 0.707 | 0.637 | 10.11 |
| wandering | stimulated | 0.236 | 0.1928 | 1.224 | 0.2345 | 0.707 | 0.637 | 10.11 |
| flow | dancing | 0.0678 | 0.2067 | 0.3282 | 0.746 | 0.721 | 0.654 | 10.83 |
| force | dancing | -0.6624 | 0.2832 | -2.3385 | 0.0293 | 0.721 | 0.654 | 10.83 |
| interior | dancing | 0.8729 | 0.2848 | 3.0655 | 0.0059 | 0.721 | 0.654 | 10.83 |
| movement | dancing | 0.2233 | 0.3142 | 0.7108 | 0.485 | 0.721 | 0.654 | 10.83 |
| wandering | dancing | 0.323 | 0.1881 | 1.7166 | 0.1008 | 0.721 | 0.654 | 10.83 |
| flow | entertained | 0.0523 | 0.1483 | 0.353 | 0.7276 | 0.856 | 0.822 | 25.02 |
| force | entertained | 0.6329 | 0.2032 | 3.1151 | 0.0052 | 0.856 | 0.822 | 25.02 |
| interior | entertained | -0.2424 | 0.2043 | -1.1869 | 0.2485 | 0.856 | 0.822 | 25.02 |
| movement | entertained | 0.298 | 0.2253 | 1.3222 | 0.2003 | 0.856 | 0.822 | 25.02 |
| wandering | entertained | 0.2617 | 0.135 | 1.9393 | 0.066 | 0.856 | 0.822 | 25.02 |
| flow | energized | 0.1889 | 0.2252 | 0.8389 | 0.411 | 0.669 | 0.59 | 8.47 |
| force | energized | -0.4009 | 0.3085 | -1.2993 | 0.2079 | 0.669 | 0.59 | 8.47 |
| interior | energized | 0.5632 | 0.3102 | 1.8156 | 0.0837 | 0.669 | 0.59 | 8.47 |
| movement | energized | 0.0699 | 0.3422 | 0.2041 | 0.8402 | 0.669 | 0.59 | 8.47 |
| wandering | energized | 0.446 | 0.2049 | 2.1764 | 0.0411 | 0.669 | 0.59 | 8.47 |
| flow | moving | 0.1182 | 0.1765 | 0.6695 | 0.5104 | 0.796 | 0.748 | 16.43 |
| force | moving | 0.4384 | 0.2418 | 1.8133 | 0.0841 | 0.796 | 0.748 | 16.43 |
| interior | moving | 0.0222 | 0.2431 | 0.0911 | 0.9282 | 0.796 | 0.748 | 16.43 |
| movement | moving | 0.2078 | 0.2682 | 0.775 | 0.447 | 0.796 | 0.748 | 16.43 |
| wandering | moving | 0.198 | 0.1606 | 1.2327 | 0.2313 | 0.796 | 0.748 | 16.43 |
| flow | animated | 0.0355 | 0.1759 | 0.2018 | 0.842 | 0.798 | 0.75 | 16.57 |
| force | animated | -0.49 | 0.241 | -2.0331 | 0.0549 | 0.798 | 0.75 | 16.57 |
| interior | animated | 0.6154 | 0.2423 | 2.5401 | 0.019 | 0.798 | 0.75 | 16.57 |
| movement | animated | 0.5422 | 0.2673 | 2.0284 | 0.0554 | 0.798 | 0.75 | 16.57 |
| wandering | animated | 0.2023 | 0.1601 | 1.2641 | 0.22 | 0.798 | 0.75 | 16.57 |
| flow | excited | -0.161 | 0.1535 | -1.0485 | 0.3063 | 0.846 | 0.809 | 23.06 |
| force | excited | -0.3154 | 0.2103 | -1.4996 | 0.1486 | 0.846 | 0.809 | 23.06 |
| interior | excited | 0.1864 | 0.2115 | 0.8813 | 0.3881 | 0.846 | 0.809 | 23.06 |
| movement | excited | 0.758 | 0.2333 | 3.2492 | 0.0038 | 0.846 | 0.809 | 23.06 |
| wandering | excited | 0.4321 | 0.1397 | 3.0925 | 0.0055 | 0.846 | 0.809 | 23.06 |
| flow | rhythm | 0.3157 | 0.2145 | 1.472 | 0.1558 | 0.699 | 0.628 | 9.768 |
| force | rhythm | 0.093 | 0.2939 | 0.3166 | 0.7547 | 0.699 | 0.628 | 9.768 |
| interior | rhythm | 0.3707 | 0.2954 | 1.2547 | 0.2234 | 0.699 | 0.628 | 9.768 |
| movement | rhythm | 0.3336 | 0.3259 | 1.0235 | 0.3177 | 0.699 | 0.628 | 9.768 |
| wandering | rhythm | -0.3046 | 0.1952 | -1.5604 | 0.1336 | 0.699 | 0.628 | 9.768 |

Table 3b. Multiple Linear Regression Result of Imagery and Entertainment from Tailors

Table 3b shows the multiple linear regression analysis results between five music imagery aspects and eight music entertainment features in timbre visualization. Red letters in each table represent a significant positive predictor variable with a predicted value. Blue letters in each table represent a significant negative predictor variable with a predicted value. Except for the moving ($F(5, 21)=16.34$, $p>0.05$) and rhythm ($F(5, 21)=9.7680$, $p>0.05$) aspect, which had no significant timbral feature, I was able to observe significant effects for the last six music entertainment aspects. For the stimulated aspects, it was found to be imagery feature of the interior turned out to be a significant and positive leading cause ($F(5, 21)=10.11$, $p=0.0068$). Also, the imagery feature of the interior turned out to be the main cause of the entertainment feature of dancing ($F(5, 21)=10.83$, $p=0.0059$), and the feature of animated ($F(5, 21)=16.57$, $p=0.019$) too. The regression coefficient of the interior was 0.8757 for the entertainment of stimulated ($t(21)=3.0006$, $p=0.0068$), 0.8729 for the entertainment of dancing ($t(21)=3.0655$, $p=0.0059$), and 0.6154 for the entertainment of animated ($t(21)=2.5401$, $p=0.0173$) respectively. For the entertaining aspect of entertainment, the imagery feature of force was found to have a strong, positive correlation ($F(5, 21)=25.02$, $p=0.0052$), showing a positive correlation with coefficients 0.6239 ($t(21)=3.1151$, $p=0.0052$). In the energized aspect of entertainment, the imagery feature of wandering was found to have a weak positive correlation ($F(5, 21)=8.47$, $p=0.0411$), showing a coefficient of 0.4460 ($t(21)=2.1764$, $p=0.0411$). For the exciting aspect of entertainment, the imagery feature of movement ($F(5, 21)=23.06$, $p=0.0038$) and wandering ($F(5, 21)=23.06$, $p=0.0055$) was found to have positive correlations. Coefficients were found to be 0.7580 and 0.4321 each, which can be interpreted there was a strong positive correlation between movement and wandering ($t(21)=3.2492$, $p=0.0038$), and a weak positive correlation between wandering and excited ($t(21)=3.0925$, $p=0.0055$).

Consequently, except for the moving and rhythm aspect of entertainment, users were affected by their internalization of music, to entertain the music in the way of feeling shuddered (stimulated), willing to dance (dancing), and a sense of liveliness (animated). For entertaining the music (entertained), the musical force that the user got was a plain positive factor. The imagery of wandering around (wandering) was a positive factor for energized feeling (energized). And lastly, for the exciting feeling (excited), the imagery of movement and wandering around was a positive cause.

#### 4. 2. 4. Coefficients Comparison by Fisher Transformation

| Condition | IV | DV | coefficients | Condition | IV | DV | coefficients | p-value |
|---|---|---|---|---|---|---|---|---|
| A Only Music | hard | flow | -0.103 | C Timbre Visualization | hard | flow | 0.2738 | |
| | soft | | 0.0386 | | soft | | -0.0493 | |
| | deep | | 0.0429 | | deep | | -0.1238 | |
| | shallow | | 0.0745 | | shallow | | 0.1457 | |
| | **bright** | | **0.9809** | | **bright** | | **-0.0587** | *** |
| | dark | | 0.8081 | | dark | | -0.6351 | |
| | warm | | 0.0863 | | warm | | 0.3726 | |
| | cold | | 0.0392 | | cold | | 0.3282 | |
| | **rough** | | **-0.257** | | **rough** | | **0.2802** | * |
| | smooth | | -0.1674 | | smooth | | 0.1690 | |
| | sharp | | -0.3402 | | sharp | | 0.2765 | |
| | **blunt** | | **-0.7973** | | **blunt** | | **0.1497** | *** |
| | **hard** | force | **-0.1789** | | **hard** | force | **0.4779** | *** |
| | soft | | 0.2397 | | soft | | 0.0455 | |
| | deep | | -0.5186 | | deep | | -0.5263 | |
| | shallow | | -0.4125 | | shallow | | -0.0252 | |
| | **bright** | | **0.6853** | | **bright** | | **0.2806** | * |
| | **dark** | | **0.6534** | | **dark** | | **-0.2563** | *** |
| | warm | | 0.1972 | | warm | | 0.1611 | |
| | cold | | 0.0755 | | cold | | 0.4237 | |
| | rough | | 0.3003 | | rough | | 0.4832 | |
| | smooth | | 0.4922 | | smooth | | 0.2476 | |
| | sharp | | -0.0888 | | sharp | | 0.0257 | |
| | blunt | | -0.7227 | | blunt | | -0.0441 | |
| | **hard** | interior | **-0.153** | | **hard** | interior | **0.5569** | ** |
| | soft | | 0.2342 | | soft | | 0.2463 | |
| | deep | | -0.4938 | | deep | | -0.4004 | |
| | shallow | | -0.51 | | shallow | | -0.3034 | |
| | bright | | 0.4905 | | bright | | 0.1443 | |
| | **dark** | | **0.6202** | | **dark** | | **-0.2920** | *** |
| | warm | | 0.2809 | | warm | | 0.0647 | |
| | **cold** | | **0.0377** | | **cold** | | **0.7377** | *** |
| | rough | | 0.3299 | | rough | | 0.6670 | |
| | smooth | | 0.4347 | | smooth | | 0.4860 | |
| | sharp | | -0.0564 | | sharp | | -0.2274 | |
| | blunt | | -0.6837 | | blunt | | -0.5799 | |
| | **hard** | movement | **-0.1804** | | **hard** | movement | **0.6692** | *** |
| | soft | | 0.2765 | | soft | | 0.2058 | |
| | deep | | -0.0577 | | deep | | -0.1190 | |
| | shallow | | 0.0525 | | shallow | | 0.2842 | |
| | bright | | 0.2852 | | bright | | 0.0078 | |
| | dark | | -0.084 | | dark | | -0.4926 | |
| | warm | | 0.2389 | | warm | | -0.0814 | |
| | **cold** | | **0.1339** | | **cold** | | **0.5876** | * |
| | **rough** | | **-0.1139** | | **rough** | | **0.6651** | ** |
| | smooth | | 0.0508 | | smooth | | 0.1293 | |
| | **sharp** | | **0.2184** | | **sharp** | | **-0.4375** | ** |
| | **blunt** | | **0.0104** | | **blunt** | | **-0.4756** | * |
| | **hard** | wandering | **-0.5014** | | **hard** | wandering | **0.1795** | ** |
| | soft | | 0.0325 | | soft | | 0.1320 | |
| | deep | | -0.0694 | | deep | | 0.0479 | |
| | shallow | | -0.2037 | | shallow | | -0.2254 | |
| | bright | | -0.0709 | | bright | | -0.1087 | |
| | **dark** | | **0.2794** | | **dark** | | **-0.3357** | * |
| | warm | | 0.0752 | | warm | | 0.0725 | |
| | **cold** | | **-0.3055** | | **cold** | | **0.3854** | ** |
| | **rough** | | **0.2961** | | **rough** | | **0.7479** | * |
| | smooth | | 0.248 | | smooth | | 0.4513 | |
| | **sharp** | | **0.9135** | | **sharp** | | **0.0948** | * |
| | blunt | | 0.2217 | | blunt | | -0.1235 | |

Table 4a. Fisher Transformation Result for Timbre → Imagery (A vs. C)

| Condition | IV | DV | coefficients | Condition | IV | DV | coefficients | p-value |
|---|---|---|---|---|---|---|---|---|
| B Basic Visualization | hard | flow | 0.4495 | C Timbre Visualization | hard | flow | 0.2738 | |
| | soft | | 0.105 | | soft | | -0.0493 | |
| | deep | | -0.3322 | | deep | | -0.1238 | |
| | shallow | | 0.2961 | | shallow | | 0.1457 | |
| | bright | | 0.1752 | | bright | | -0.0587 | |
| | **dark** | | **0.1782** | | **dark** | | **-0.6351** | *** |
| | warm | | -0.0088 | | warm | | 0.3726 | |
| | cold | | -0.1185 | | cold | | 0.3282 | |
| | rough | | 0.2853 | | rough | | 0.2802 | |
| | smooth | | -0.2745 | | smooth | | 0.1690 | |
| | sharp | | 0.325 | | sharp | | 0.2765 | |
| | blunt | | 0.2194 | | blunt | | 0.1497 | |
| | hard | force | 0.3668 | | hard | force | 0.4779 | |
| | soft | | 0.0997 | | soft | | 0.0455 | |
| | deep | | -0.5839 | | deep | | -0.5263 | |
| | shallow | | -0.0195 | | shallow | | -0.0252 | |
| | bright | | 0.0446 | | bright | | 0.2806 | |
| | dark | | 0.0625 | | dark | | -0.2563 | |
| | warm | | 0.175 | | warm | | 0.1611 | |
| | cold | | 0.0246 | | cold | | 0.4237 | |
| | rough | | 0.4262 | | rough | | 0.4832 | |
| | smooth | | 0.1198 | | smooth | | 0.2476 | |
| | sharp | | 0.2265 | | sharp | | 0.0257 | |
| | blunt | | 0.0685 | | blunt | | -0.0441 | |
| | hard | interior | 0.2559 | | hard | interior | 0.5569 | |
| | soft | | 0.148 | | soft | | 0.2463 | |
| | deep | | -0.4 | | deep | | -0.4004 | |
| | shallow | | -0.1763 | | shallow | | -0.3034 | |
| | bright | | 0.0267 | | bright | | 0.1443 | |
| | **dark** | | **0.1858** | | **dark** | | **-0.2920** | * |
| | warm | | 0.3592 | | warm | | 0.0647 | |
| | **cold** | | **0.1422** | | **cold** | | **0.7377** | ** |
| | **rough** | | **0.1036** | | **rough** | | **0.6670** | ** |
| | smooth | | 0.0917 | | smooth | | 0.4860 | |
| | **sharp** | | **0.2896** | | **sharp** | | **-0.2274** | * |
| | **blunt** | | **-0.0432** | | **blunt** | | **-0.5799** | * |
| | **hard** | movement | **0.1978** | | **hard** | movement | **0.6692** | * |
| | soft | | 0.1233 | | soft | | 0.2058 | |
| | deep | | -0.1655 | | deep | | -0.1190 | |
| | **shallow** | | **0.7099** | | **shallow** | | **0.2842** | * |
| | **bright** | | **0.4993** | | **bright** | | **0.0078** | * |
| | **dark** | | **0.2309** | | **dark** | | **-0.4926** | ** |
| | warm | | 0.0085 | | warm | | -0.0814 | |
| | **cold** | | **-0.1075** | | **cold** | | **0.5876** | ** |
| | **rough** | | **-0.2333** | | **rough** | | **0.6651** | *** |
| | **smooth** | | **-0.6251** | | **smooth** | | **0.1293** | ** |
| | **sharp** | | **0.0463** | | **sharp** | | **-0.4375** | * |
| | blunt | | -0.0823 | | blunt | | -0.4756 | |
| | hard | wandering | -0.1656 | | hard | wandering | 0.1795 | |
| | soft | | -0.2036 | | soft | | 0.1320 | |
| | deep | | 0.0854 | | deep | | 0.0479 | |
| | **shallow** | | **0.2774** | | **shallow** | | **-0.2254** | * |
| | **bright** | | **0.3554** | | **bright** | | **-0.1087** | * |
| | **dark** | | **0.2487** | | **dark** | | **-0.3357** | * |
| | warm | | 0.3929 | | warm | | 0.0725 | |
| | **cold** | | **-0.407** | | **cold** | | **0.3854** | ** |
| | **rough** | | **-0.0543** | | **rough** | | **0.7479** | *** |
| | **smooth** | | **-0.3599** | | **smooth** | | **0.4513** | ** |
| | **sharp** | | **0.5396** | | **sharp** | | **0.0948** | * |
| | blunt | | 0.1753 | | blunt | | -0.1235 | |

Table 4b. Fisher Transformation Result for Timbre → Imagery (B vs. C)

Fisher transformation was used to compare coefficients and determine if Tailors effectively conveyed timbral features, made music imagery, and made users to entertained the music among the three conditions. P-values were also evaluated to determine whether the coefficients' difference was significant. Each aestrisk notations in the table represents pvalue ranges: * p<0.05, ** p<0.01, ** *p<0.001. Table 4a represents the comparison results between coefficients in multiple regression equations (Timbre → Imagery) from comparing condition A (Only Music) and condition C (Tailors). I found that timbral hardness significantly affected more to imagery features of force(r=0.47), interior(r=0.55), and movement(r=0.66) in Tailors than in only music. Also, timbral coldness significantly affected more to imagery features of the interior(r=0.73) and movement(r=0.58) in Tailors. Lastly, timbral roughness significantly affected imagery features of movement(r=0.66) and wandering(r=0.74) higher in Tailors too. This result shows that, except for imagery features of flow, timbral hardness, coldness, and roughness of Tailors were significantly higher predictors for users to make last music imageries. However, there were timbral features that did not work well as only music condition. Firstly, timbral brightness was a significantly lower predictor for flow(r=-0.05) and force(r=0.28) imagery features in Tailors than only music. Timbral darkness was also found to be a significantly lower predictor for the imagery feature of force(r=-0.25). Lastly, timbral sharpness was a significantly lower predictor for the imagery feature of wandering(r=0.09). This result shows that timbre visualization was weaker than the only music condition in these timbral features.

Table 4b shows the comparison results between coefficients in multiple regression equations from comparing condition B (Basic Visualization) and condition C (Tailors). Timbral roughness significantly affected imagery features of the interior(r=0.66), movement(r=0.66), and wandering(r=0.74) higher on Tailors than basic visualization. Timbral coldness affected significantly more in Tailors to imagery features of the interior(r=0.73) and movement(r=0.58). Also, timbral hardness is affected more in timbre visualization to imagery features of movement(r=0.66). Lastly, timbral smoothness was affected more in imagery features of wandering(r=0.45). It shows that except for the imagery feature of flow, timbral roughness, coldness, hardness, and smoothness were significantly more effective in making music imagery in timbre visualization than in basic visualization. On the contrary, imagery features of the movement were significantly less affected by timbral brightness(r=0.00), darkness(r=-0.49), sharpness(r=-0.43), and shallowness(r=0.28). Also, timbral darkness was a significantly lower predictor for the imagery feature of flow(r=-0.63). Lastly, timbral sharpness was a significantly lower predictor for the imagery feature of wandering(r=0.09). This result shows timbre visualization was weaker than the basic visualization in these timbral features.

| Condition | IV | DV | coefficients | Condition | IV | DV | coefficients | p-value |
|---|---|---|---|---|---|---|---|---|
| **A Only Music** | flow | **stimulated** | 0.1962 | **C Timbre Visualization** | flow | **stimulated** | -0.0453 | |
| | force | | -0.5755 | | force | | -0.7460 | |
| | **interior** | | **0.4723** | | **interior** | | **0.8757** | ** |
| | movement | | 0.0656 | | movement | | 0.4512 | |
| | **wandering** | | **0.7363** | | **wandering** | | **0.2360** | ** |
| | flow | **dancing** | -0.0523 | | flow | **dancing** | 0.0678 | |
| | **force** | | **0.0730** | | **force** | | **-0.6624** | ** |
| | **interior** | | **0.3238** | | **interior** | | **0.8729** | *** |
| | movement | | 0.3520 | | movement | | 0.2233 | |
| | wandering | | 0.2739 | | wandering | | 0.3230 | |
| | flow | **entertained** | 0.0456 | | flow | **entertained** | 0.0523 | |
| | **force** | | **1.0340** | | **force** | | **0.6329** | *** |
| | interior | | -0.2334 | | interior | | -0.2424 | |
| | movement | | -0.0795 | | movement | | 0.2980 | |
| | wandering | | 0.1800 | | wandering | | 0.2617 | |
| | flow | **energized** | 0.2932 | | flow | **energized** | 0.1889 | |
| | **force** | | **0.4794** | | **force** | | **-0.4009** | *** |
| | **interior** | | **-0.0348** | | **interior** | | **0.5632** | ** |
| | movement | | -0.3496 | | movement | | 0.0699 | |
| | wandering | | 0.4829 | | wandering | | 0.4460 | |
| | flow | **moving** | -0.0533 | | flow | **moving** | 0.1182 | |
| | force | | 0.4629 | | force | | 0.4384 | |
| | **interior** | | **0.5392** | | **interior** | | **0.0222** | * |
| | movement | | -0.0066 | | movement | | 0.2078 | |
| | wandering | | 0.0173 | | wandering | | 0.1980 | |
| | flow | **animated** | 0.3358 | | flow | **animated** | 0.0355 | |
| | force | | -0.5121 | | force | | -0.4900 | |
| | interior | | 0.4409 | | interior | | 0.6154 | |
| | movement | | 0.3145 | | movement | | 0.5422 | |
| | wandering | | 0.4264 | | wandering | | 0.2023 | |
| | flow | | 0.1020 | | flow | | -0.1610 | |
| | force | | -0.2929 | | force | | -0.3154 | |
| | interior | **excited** | 0.4931 | | interior | **excited** | 0.1864 | |
| | **movement** | | **0.1575** | | **movement** | | **0.7580** | ** |
| | wandering | | 0.5602 | | wandering | | 0.4321 | |
| | flow | | 0.5216 | | flow | | 0.3157 | |
| | force | | -0.0033 | | force | | 0.0930 | |
| | interior | **rhythm** | 0.1679 | | interior | **rhythm** | 0.3707 | |
| | movement | | 0.3603 | | movement | | 0.3336 | |
| | wandering | | -0.2700 | | wandering | | -0.3046 | |

Table 5a. Fisher Transformation Result for Imagery → Entertain (A vs. C)

| Condition | IV | DV | coefficients | Condition | IV | DV | coefficients | p-value |
|---|---|---|---|---|---|---|---|---|
| | flow | | -0.2220 | | flow | | -0.0453 | |
| | **force** | | **-0.2313** | | **force** | | **-0.7460** | ** |
| | **interior** | stimulated | **0.4779** | | **interior** | stimulated | **0.8757** | ** |
| | movement | | 0.1413 | | movement | | 0.4512 | |
| | wandering | | 0.5000 | | wandering | | 0.2360 | |
| | flow | | -0.2510 | | flow | | 0.0678 | |
| | force | | -0.4700 | | force | | -0.6624 | |
| | **interior** | dancing | **0.6344** | | **interior** | dancing | **0.8729** | * |
| | movement | | 0.4516 | | movement | | 0.2233 | |
| | wandering | | 0.3631 | | wandering | | 0.3230 | |
| | flow | | -0.0182 | | flow | | 0.0523 | |
| | force | | 0.7845 | | force | | 0.6329 | |
| | **interior** | entertained | **-0.7951** | | **interior** | entertained | **-0.2424** | ** |
| B Basic Visualization | movement | | 0.4078 | C Timbre Visualization | movement | | 0.2980 | |
| | wandering | | 0.3732 | | wandering | | 0.2617 | |
| | flow | | 0.0701 | | flow | | 0.1889 | |
| | **force** | | **0.0926** | | **force** | | **-0.4009** | * |
| | interior | energized | 0.2257 | | interior | energized | 0.5632 | |
| | movement | | -0.0930 | | movement | | 0.0699 | |
| | wandering | | 0.5849 | | wandering | | 0.4460 | |
| | flow | | 0.1667 | | flow | | 0.1182 | |
| | force | | 0.1753 | | force | | 0.4384 | |
| | **interior** | moving | **0.4829** | | **interior** | moving | **0.0222** | * |
| | movement | | 0.0755 | | movement | | 0.2078 | |
| | wandering | | 0.0183 | | wandering | | 0.1980 | |
| | flow | | 0.0616 | | flow | | 0.0355 | |
| | **force** | animated | **0.1157** | | **force** | animated | **-0.4900** | * |
| | interior | | 0.5812 | | interior | | 0.6154 | |
| | **movement** | | **-0.0621** | | **movement** | | **0.5422** | * |
| | wandering | | 0.2733 | | wandering | | 0.2023 | |
| | flow | **excited** | -0.2510 | | flow | **excited** | -0.1610 | |
| | force | | 0.0949 | | force | | -0.3154 | |
| | interior | | 0.1222 | | interior | | 0.1864 | |
| | **movement** | | **0.3437** | | **movement** | | **0.7580** | * |
| | wandering | | 0.5334 | | wandering | | 0.4321 | |
| | flow | **rhythm** | 0.6573 | | flow | **rhythm** | 0.3157 | |
| | force | | 0.1636 | | force | | 0.0930 | |
| | interior | | 0.1248 | | interior | | 0.3707 | |
| | movement | | 0.0076 | | movement | | 0.3336 | |
| | wandering | | -0.1683 | | wandering | | -0.3046 | |

Table 5b. Fisher Transformation Result for Imagery → Entertain (B vs. C)

Table 5a represents the comparison results between coefficients in multiple regression equations (Imagery → Entertainment) from comparing condition A (Only Music) and condition C (Tailors). Especially, coefficients between imagery feature of interior and the entertainment features of stimulated(r=0.87), dancing(r=0.87), and energized(r=0.56) were significantly higher in Tailors than the only music. Also, being able to move the body to the rhythm (movement) has significantly affected higher to the excited feeling(r=0.75) of users in Tailors. However, the imagery feature of wandering has less affected stimulated(r=0.23) feeling significant. Furthermore, imagery features of force have less affected feelings for dance(r=-0.66) significantly. Finally, interior imagery has less affected the entertainment feature of moving(r=0.02) significantly.

Table 5b shows the comparison results between coefficients in multiple regression equations from comparing condition B (Basic Visualization) and condition C (Tailors). The result showed that the imagery of becoming one with the music (interior) affected music entertainment more in timbre visualization than the basic visualization. Except for entertaining and moving features, the interior element of music imagery influenced more music entertainment than other music imagery features in Tailors. Especially, the coefficients between interior imagery with stimulated feeling(r=0.87) and feeling to dance(r=0.87) were significantly higher in Tailors than the basic visualization. The coefficients between movement imagery with animated(r=0.54) and excited(r=0.75) feelings were significantly higher too. On the other hand, the imagery feature of force was significantly less affected by music entertainment of stimulated(r=-0.74), energized(r=-0.40), and animated(-0.49). As same with the result in Table 5a, interior imagery has less affected the entertainment feature of moving(0.02) significantly, showing that each imagery feature can affect or not affect entertainment features due to its characteristics.

4. 3. Post Survey Results
This section is organized and based on the post-survey results (See Appendix E for the questionnaires) and found some insights from users' interviews.

4. 3. 1. Rankings for the Best Timbre Expression

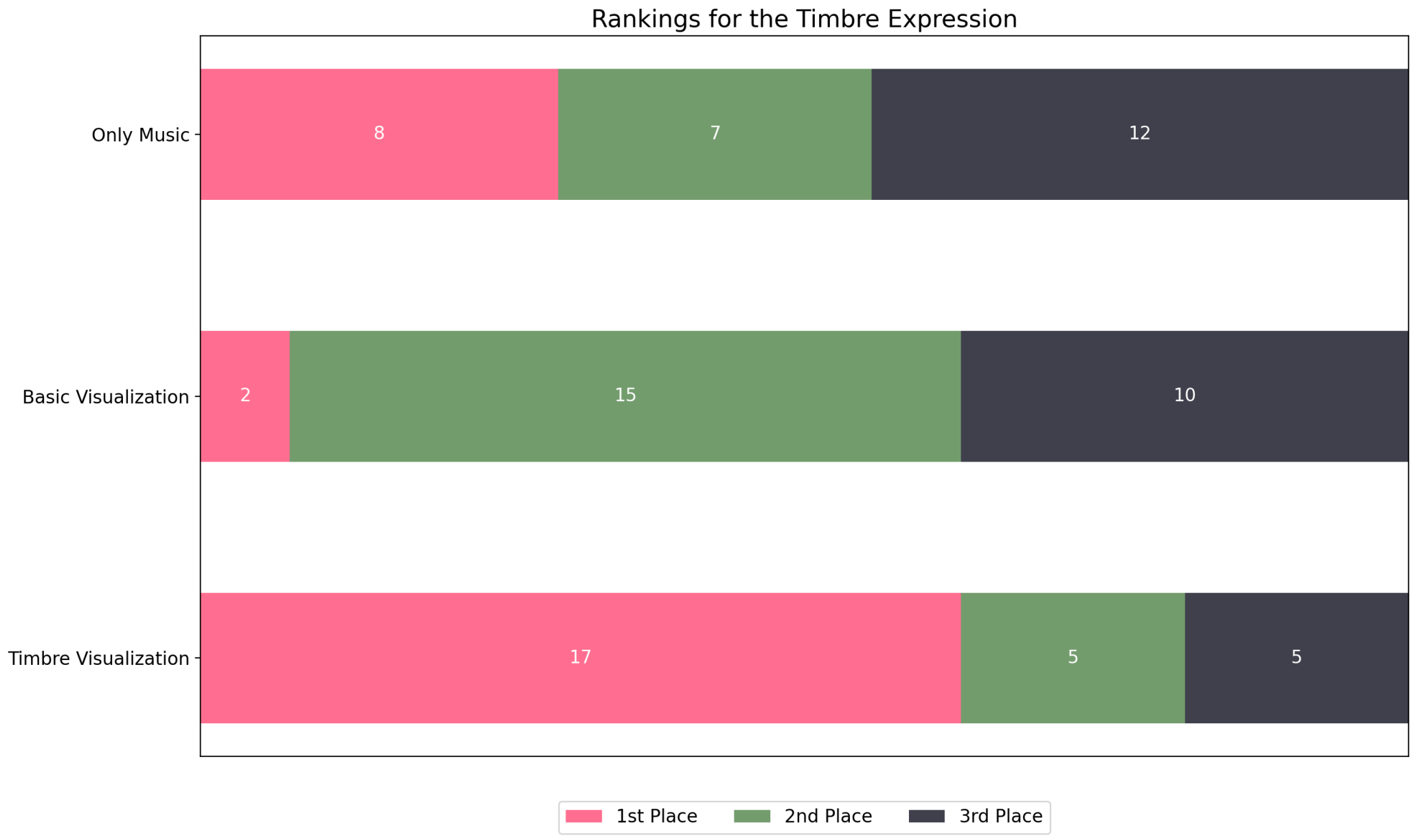

Figure 7. Ranking Comparisons for the Best Timbre Expression

Figure 7 represents that Timbre visualization made by Tailors the most well-expressed way of timbre among the three conditions. 62.96% of participants (17 out of 27) gave Tailors the first place for the rank in timbre expression. Eight people for only music condition and two for the basic visualization. The participants' opinion who gave Tailors the first place in the best timbre expression ranking was as follows; P17 said Tailors' visual background well delivered the timbral features of the music, which he thought was also related to the mood of the music. P25 stated that she gave Tailors the first place in timbre expression ranking because even the tiny sounds that users might miss were expressed in timbre visualization, which led her to enjoy and have more fun when listening to music. P24 said that she could focus more on visualization and the music itself because there were more dynamic movements and various colors. Also, P2 evaluated that Tailors' visualization immersed her immediately in the music. On the contrary, those who gave Tailors the third place were the most minor (5 out of 27); they all pointed out that timbre visualization felt dizzy because there were so many components, distracting them from focusing on music.

### 4. 3. 2. Rankings for the Best Music Experience

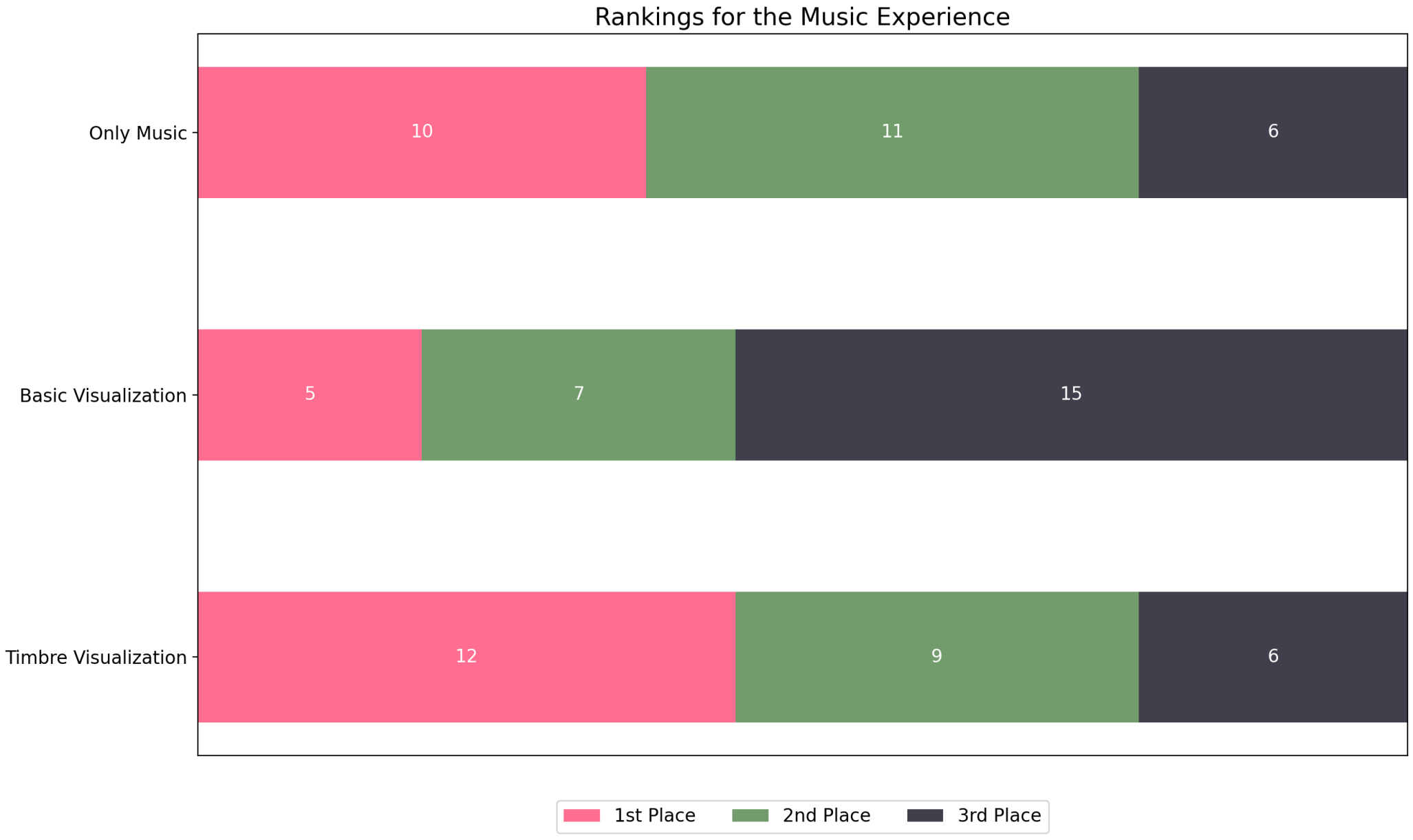

Figure 8. Ranking Comparisons for the Best Music Experience

Figure 8 represents that Tailors was the most well-enjoyable way of music listening among the three conditions. 44.44% of participants (12 out of 27) gave Tailors the first place for the rank in music experience, followed by ten people for only music condition and 5 for the basic visualization. The participants' opinion who gave Tailors the first place in the best music experience ranking was as follows: P3 stated that timbre visualization expressed the detailed background sound of the music well, making her listen to listening music more closely. Also, P19 evaluated the background with color clouds, and the sky felt well harmonized when the music was quiet and calm. P24 found that Tailors helped to find something new in the same song by listening analytically. Those who gave Tailors third place in the music experience where the most minor (6 out of 27). However, in P26, who didn't find Tailors' strength in music listening experience, the experience was a pity because the visualization have affected her music experience in a forcing way. P5, who gave second place to Tailors, and first place to only music condition in the music experience ranking, stated that music is something to listen to when it is rest time. Still, the visualization gives a burden to keep an eye on.

4. 3. 3. Rankings for the Willingness to Use Again

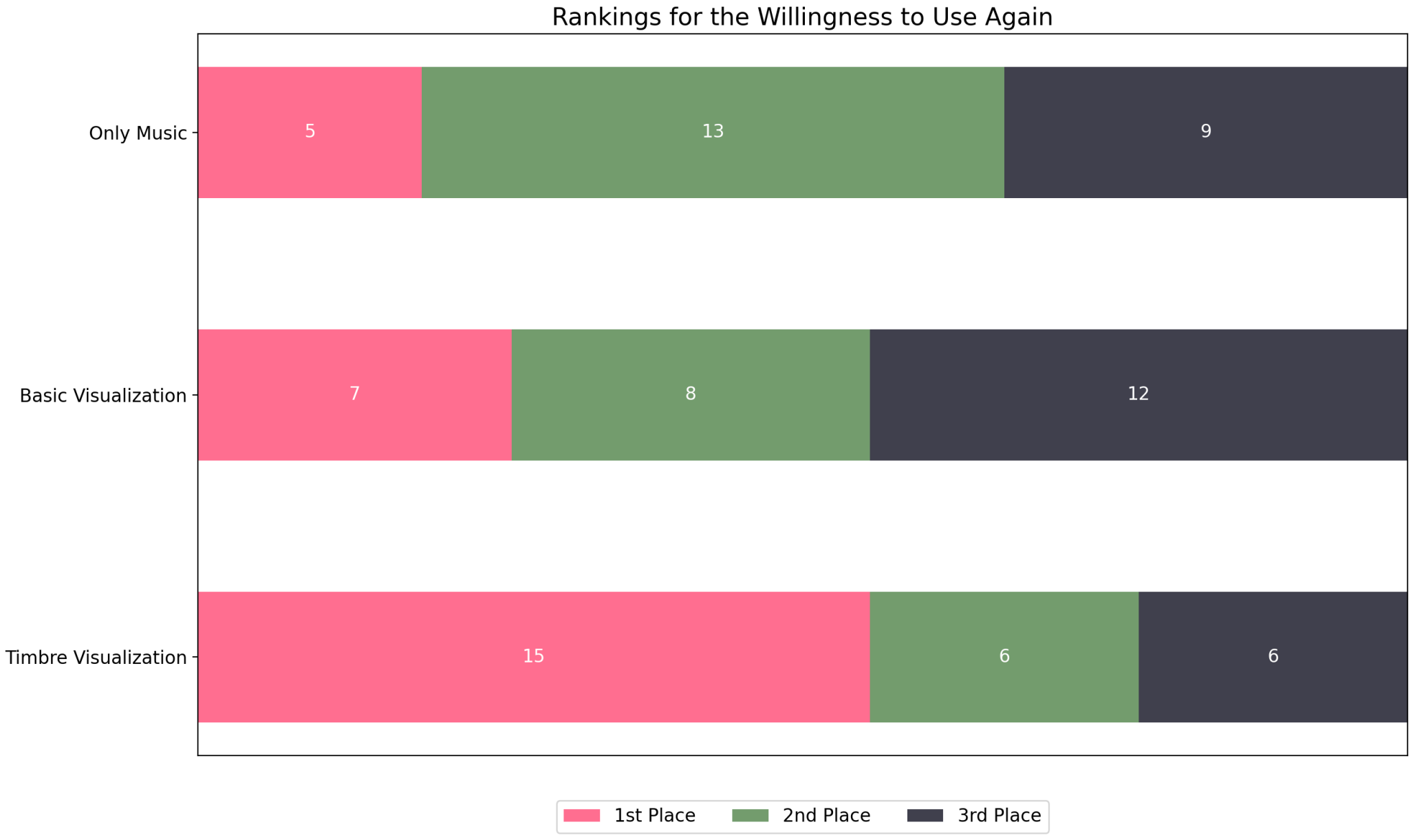

Figure 9. Ranking Comparisons for the Willingness to Use Again

Figure 9 represents that Tailors was the most reusable way of music listening among the three conditions. 55.55% of participants (15 out of 27) gave Tailors the first place for music experience, seven people for basic visualization, and 5 for the only music condition. The participants' opinion who picked Timbre visualization in the rank for use again in the first place was like these: For P16, Tailors' novelty became the main cause of picking the first rank. She picked the timbre visualization in the first place because it is a unique way to experience visualization and entertain music. P24 also stated Tailors as a fresh way to entertain music, mentioning the chance that Tailors will enable users to feel every single different timbre of the singers. P13 emphasized the feeling of falling for the music more deeply when listening to music through timbre visualization. On the other hand, those who picked Tailors' visualization for the third place (6 out of 27) found fault with the durability of watching the timbre visualization due to dizziness. There were also the participants who gave Tailors in the first place who pointed out the weak durability of watching the visualization, especially on the object front of the background. P5 pointed out that the circle objects in the front were too flashy, which was a little burdensome to experience Tailors but also had the willingness to reuse timbre visualization when the circle objects in the front were edited more clearly.

# V. DISCUSSION

5. 1. Possibilities of Tailors

Through Tailors, I found that timbre visualization affects the MIR users, which is beyond further than just conveying timbral features of the music. Each participant was able to relate each music's tone color with the timbre visualization, to shape their imagery of the music. When users hear the music, they were able to get information from both sides- musical audio and visualization. While users' experience with the music was focused on the mood and the emotion, Tailors have given opportunities to users to enjoy music by listening carefully for the timbre of the music. Also, participants anticipated their behavior of listening carefully to the music to feel the timbre although there was no notification before the experiment. By not focusing on the mood or the emotion but focusing on the timbre and the music's visual imagery, users of Tailors were able to be immersed in the timbral feature and define their subjective judgments of the mind.

From this point of the view, Tailors have made users think about various words related to the timbral features of the music. I was also able to find that semantic descriptors of timbre, which are highly related to the user's thoughts on the music, can be derived and raised from the timbre visualization users. This brings up the new possibilities of timbre visualization that its users could generate and brainstorm together about the semantic descriptors of the timbre, which can be applied differently in the visualization. There are also possibilities for users to participate in the expression of the visualization. It will allow user-centered design about the semantic descriptors of timbre. For instance, visual parameters of Tailors could be accustomed to the user's preference if customization options are opened. Relating semantic descriptors with visual parameters of Tailors, each user can take a part in music visualization and have a richer music experience. Consequently, users' participation in the timbre descriptors and applicating in the visualization could arouse more agreement of users, making users more entertained by the music with the timbre. While users do interaction with finetuning these descriptors and parameters of the visualization, this would also make users find their taste and feel of the timbral features. Users' subjective thoughts on the musical timbre would also be linked to the data-driven approach of the Tailors, which is also related to future work.

5. 2. Limitations and the Future Work

Although 7 out of 12 timbral features (warm, bright, deep, shallow, hard, rough, and sharp) were well delivered in the twenty music used in the experiment, we found that Tailors didn't convey timbral darkness and bluntness well, compared to the only music and the basic visualization. In addition, through the Kruskal-Wallis Test, there was no significant difference between the comparison of the three conditions. This shows Tailors needs to be modified to compensate for the weakness (timbral darkness and bluntness) and strengthen the strength that it already has. Also, even I've united the genre for the POP of the twenty music, more future experiments with more participants needed with various kinds of music will help to give more insights into the result by music genre and music features.

As mentioned above, participants who experienced pity music experience with Tailors pointed out the fatigue of the eye due to the dizziness. Although the participants get to experience the timbre visualization for twenty music straight during the experiment, this shows a shortage in the sustainability of the timbre visualization in the case of experiencing timbre visualization daily. Since it is a well-known fact that getting information in multimodal (both visually and by hearing) burdens more than basic hearing for music listeners, the way to reduce these burdens needs to be considered. The result shows that timbral features were well expressed, and the specific features influenced the music's visual imagery and entertainment. This makes considering a way to reduce visual components to concentrate on specific four to six timbral features rather than representing all features. Future work should focus on visually expressing timbral features that can be well delivered and the features that affect music imagery and entertainment.

Since the post-survey showed that MIR users could concentrate and be immersed in the music by timbre visualization rather than the only music and the basic visualization, future work should consider the data-driven approach to improve Tailors' mapping rules. I believe this will involve a data collection process from the MIR users to determine which timbral features connect to the elements of music visualization. Furthermore, the way of applying Tailors more artistically can be discussed. I believe using a beam projector to express timbre visualization on the wall or the ground is possible. It will help to focus on the entertainment purpose to emphasize the strength of concentrating and immersing in the timbre visualization in a specific space. Users can experience timbre visualization with the sense of the space, extending its possibility to the artistic purpose of Tailors.

# VI. CONCLUSION

In this paper, I made a music visualization system based on the timbral features called Tailors. After the experiment with 27 participants with twenty music with three different conditions, I found that Tailors effectively delivered timbral warmth, brightness, depth, shallowness, hardness, roughness, and sharpness compared to the only music condition and basic visualization. Also, five features of music imagery and eight features of music entertainment were all highest in the Tailors among the three conditions. After the multiple linear regression analysis between timbre-imagery, and imagery-entertainment, we found significant and positive correlations. This result shows that timbre visualization by Tailors made users' music visual imagery well and led to music entertainment. Post survey result of participants shows Tailors all ranked first for the best timbre expression, music experience, and willingness to try the next time again. While leaving the limitations on the usability for some users leaving a burden in the eye, Tailors led to the future work improving its mapping rule by data-driven approach from MIR users. Furthermore, focusing on the timbral and visual features of Tailors express well, in a more artistic way using the space like a plane wall or the ground, emphasizing its entertainment purpose.

# Appendix

Appendix A. User Interface of Tailors

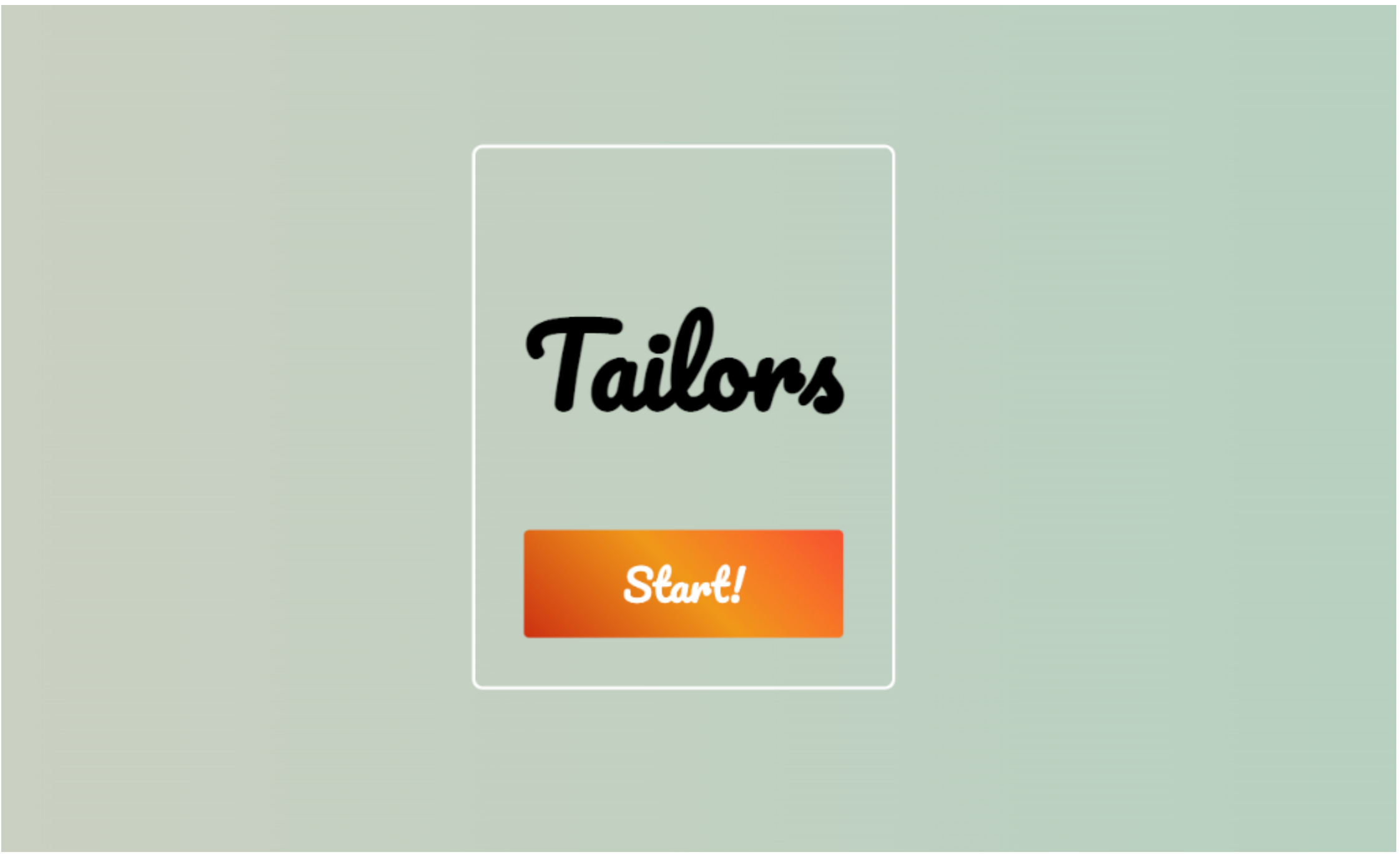

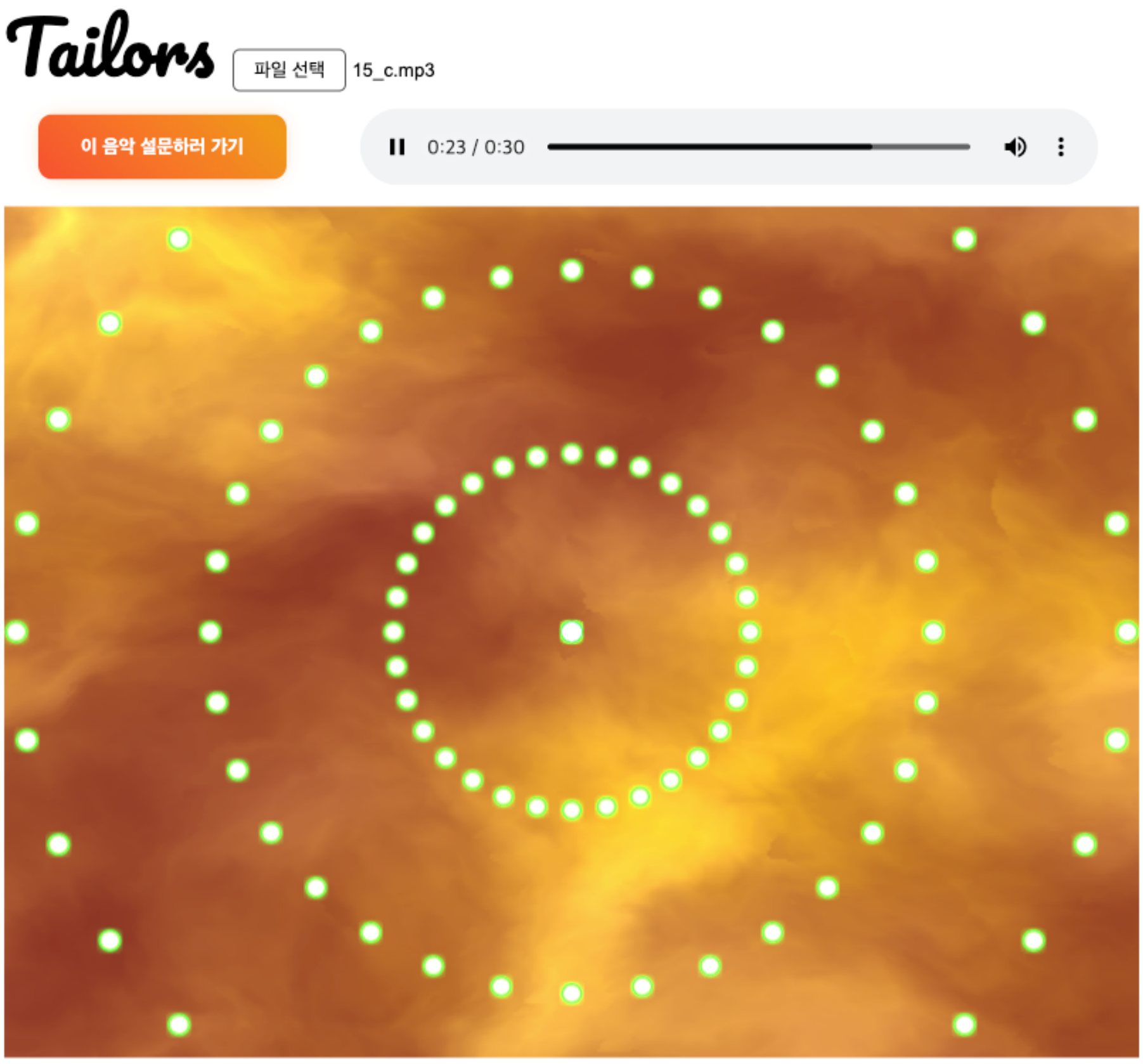

Appendix B. Information of Music Used In the Experiment

| Index | Artist | Title of the Music | Genre | Detailed Genre |
|---|---|---|---|---|
| 1 | Sufjan Stevens | Mystery of Love | POP | Folk |
| 2 | Lana Del Rey | hope is a dangerous thing for a woman like me to have - but i have it | | Alternative & Indie |
| 3 | Foster The People | Imagination | | Soul |
| 4 | Cigarette After Sex | Apocalypse | | Alternative & Indie |
| 5 | Ben Folds | Red Is Blue | | Folk |
| 6 | John Mayer | Rosie | | Folk |
| 7 | The Eurythmics | Sweet Dreams | | R&B |
| 8 | Gotye | State Of The Art | | Rock |
| 9 | Demi Lovato | Cool For The Summer | | Electronic |
| 10 | Scarlett Johansson | The Moon Song | | Alternative & Indie |
| 11 | Radiohead | Burn The Witch | | Alternative & Indie |
| 12 | Charlie Puth | That's Hilarious | | Christian |
| 13 | Ledisi | I Blame You | | R&B, Soul |
| 14 | James Vincent McMorrow | Cavalier | | Folk |
| 15 | Corinne Bailey Rae | Like A Star | | Jazz, R&B |
| 16 | Ashley Tisdale | He said She said | | R&B |
| 17 | Andrew Belle | In My Veins | | Soul |
| 18 | Ella Mai | She Don't | | R&B |
| 19 | Norah Jones | Come Away With Me | | Folk |
| 20 | Keyshia Cole | Love, I Thought You Had My Back | | R&B, Soul |

Appendix C. Visualized Output of Tailors (numbers indicate indexes for the each music)

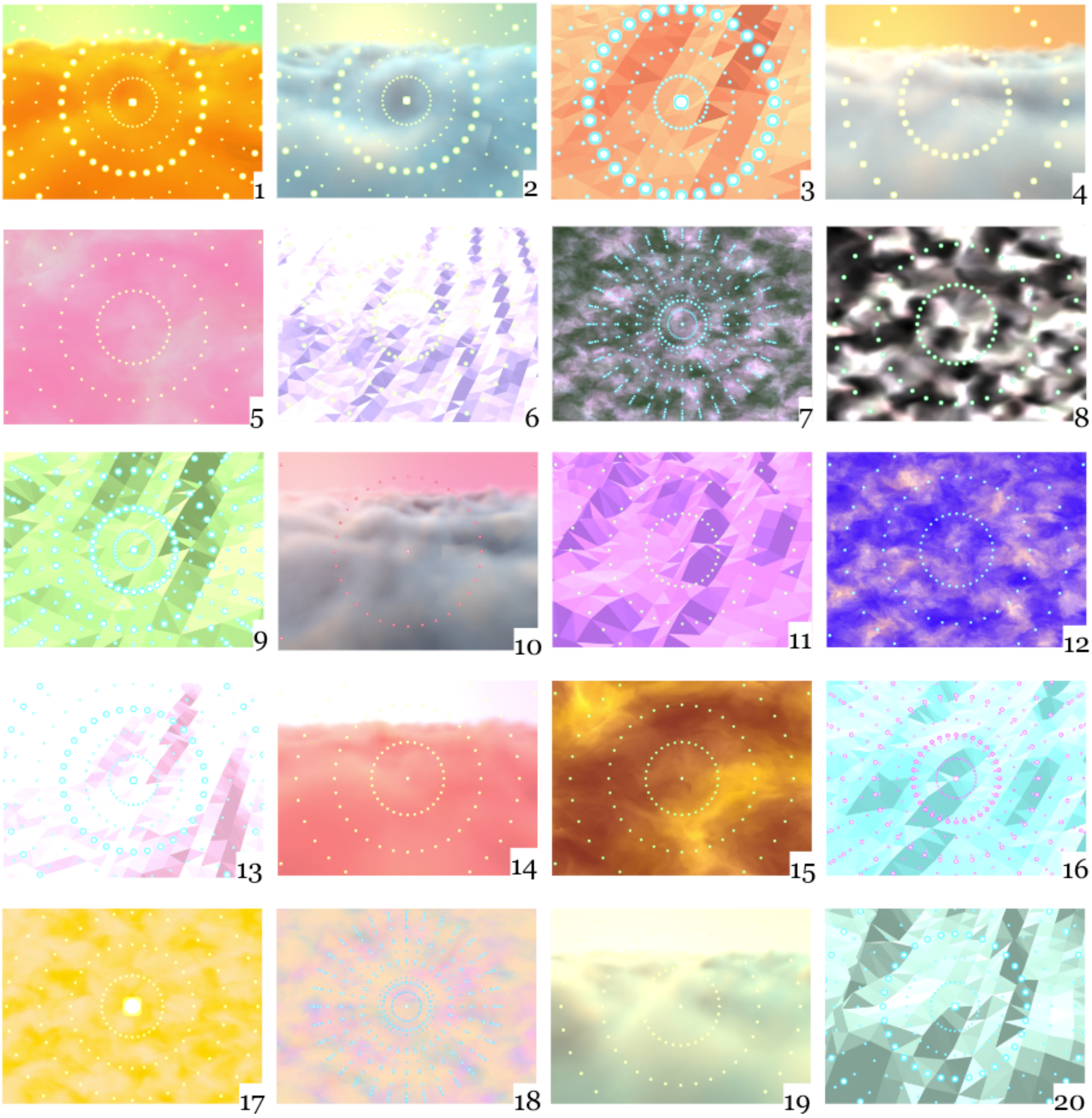

Appendix D. Tables for Wilcoxon Sign-Ranked Analysis By Surveys

## D1. Wilcoxon Sign-Ranked Analysis for the Timbre Survey

| **Significant** | **Mean Value** | | **Not Significant** | **Mean Value** | |
|---|---|---|---|---|---|
| | B | C | | B | C |
| warm (p=0.04857) | 4.1185 | **4.3333** | hard (p=0.8909) | 4.3241 | **4.3259** |
| deep(p=0.0029) | 3.3407 | **3.6333** | soft (p=0.6503) | 4.4019 | **4.4481** |
| rough (p=0.0456) | 3.9722 | **4.1593** | cold (p=0.7356) | 2.9315 | **2.9759** |
| | | | shallow(p=0.8052) | 3.7537 | **3.7833** |
| | | | blunt (p=0.2617) | **3.187** | 3.1 |
| | | | sharp (p=0.0612) | 4.5056 | **4.6667** |
| | | | smooth (p=0.4816) | 3.2296 | **3.3037** |
| | | | bright (p=0.0732) | 4.1537 | **4.3093** |
| | | | dark (p=0.2161) | **3.1167** | 3.013 |

| **Significant** | **Mean Value** | | **Not Significant** | **Mean Value** | |
|---|---|---|---|---|---|
| | A | C | | A | C |
| bright (p=0.0024) | 3.9278 | **4.3093** | hard (p=0.3936) | 4.2148 | **4.3259** |
| dark (p=0.0007) | 3.3889 | 3.013 | soft (p=0.1460) | **4.6074** | 4.4481 |
| warm (p=0.017) | 4.0796 | **4.3333** | deep (p=0.2311) | 3.4574 | **3.6333** |
| rough (p=0.0238) | 3.8648 | **4.1593** | shallow (p=0.7877) | 3.7574 | **3.7833** |
| blunt (p=0.0206) | **3.3222** | 3.1 | cold (p=0.8664) | **3.0259** | 2.9759 |
| | | | smooth (p=0.2337) | **3.4093** | 3.3037 |
| | | | sharp (p=0.1073) | 4.4074 | **4.6667** |

## D2. Wilcoxon Sign-Ranked Analysis for the Imagery Survey

| **Significant** | **Mean Value** | | **Not Significant** | **Mean Value** | |
|---|---|---|---|---|---|
| | B | C | | B | C |
| flow (p=0.0013) | 4.3907 | **4.7222** | force (p=0.1212) | 4.2611 | **4.4111** |
| wandering (p=0.0127) | 2.8907 | **3.1926** | interior (p=0.0618) | 3.9759 | **4.1519** |
| | | | movement (p=0.7005) | 4.0815 | **4.1278** |

| **Significant** | **Mean Value** | | **Not Significant** | **Mean Value** | |
|---|---|---|---|---|---|
| | A | C | | A | C |
| wandering (p=0.0023) | 2.7259 | **3.1926** | flow (p=0.1178) | 4.4926 | **4.7222** |
| | | | force (p=0.0614) | 4.0741 | **4.4111** |
| | | | interior (p=0.0997) | 3.7926 | **4.1519** |
| | | | movement(p=0.2692) | 3.9037 | **4.1278** |

D3. Wilcoxon Sign-Ranked Analysis for the Entertainment Survey

| **Significant** | **Mean Value** | | **Not Significant** | **Mean Value** | |
|---|---|---|---|---|---|
| | **B** | **C** | | **B** | **C** |
| entertained (p=0.0349) | 3.9944 | **4.2778** | stimulated (p=0.2745) | 3.0259 | **3.2963** |
| | | | dancing (p=0.8488) | 3.1148 | **3.287** |
| | | | energized (p=0.8191) | 3.4537 | **3.7944** |
| | | | moving (p=0.1506) | 4.1037 | **4.45** |
| | | | animated (p=0.4847) | 3.7463 | **4.2537** |
| | | | excited (p=0.1943) | 3.3296 | **3.7667** |
| | | | rhythm (p=0.1576) | 4.4648 | **4.9852** |

| **Significant** | **Mean Value** | | **Not Significant** | **Mean Value** | |
|---|---|---|---|---|---|
| | **A** | **C** | | **A** | **C** |
| entertained (p=0.0068) | 3.8204 | **4.2778** | stimulated (p=0.1585) | 3.0259 | **3.2963** |
| animated (p=0.0090) | 3.7463 | **4.2537** | dancing (p=0.8717) | 3.1148 | **3.287** |
| excited (p=0.0245) | 3.3296 | **3.7667** | energized (p=0.0550) | 3.4537 | **3.7944** |
| rhythm (p=0.0079) | 4.4648 | **4.9852** | moving (p=0.1331) | 4.1037 | **4.45** |

Appendix E. Post Survey Questionnaires

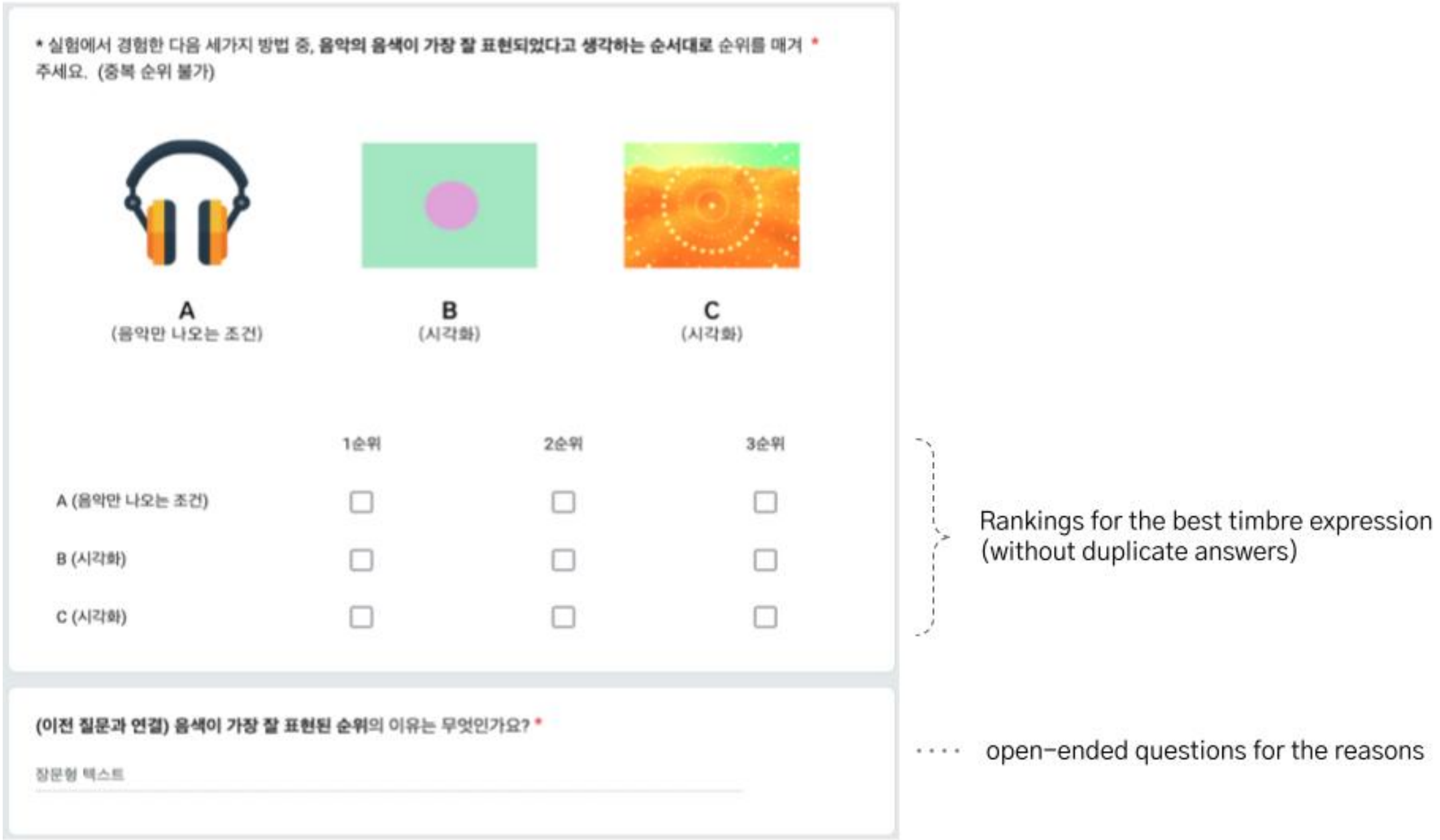

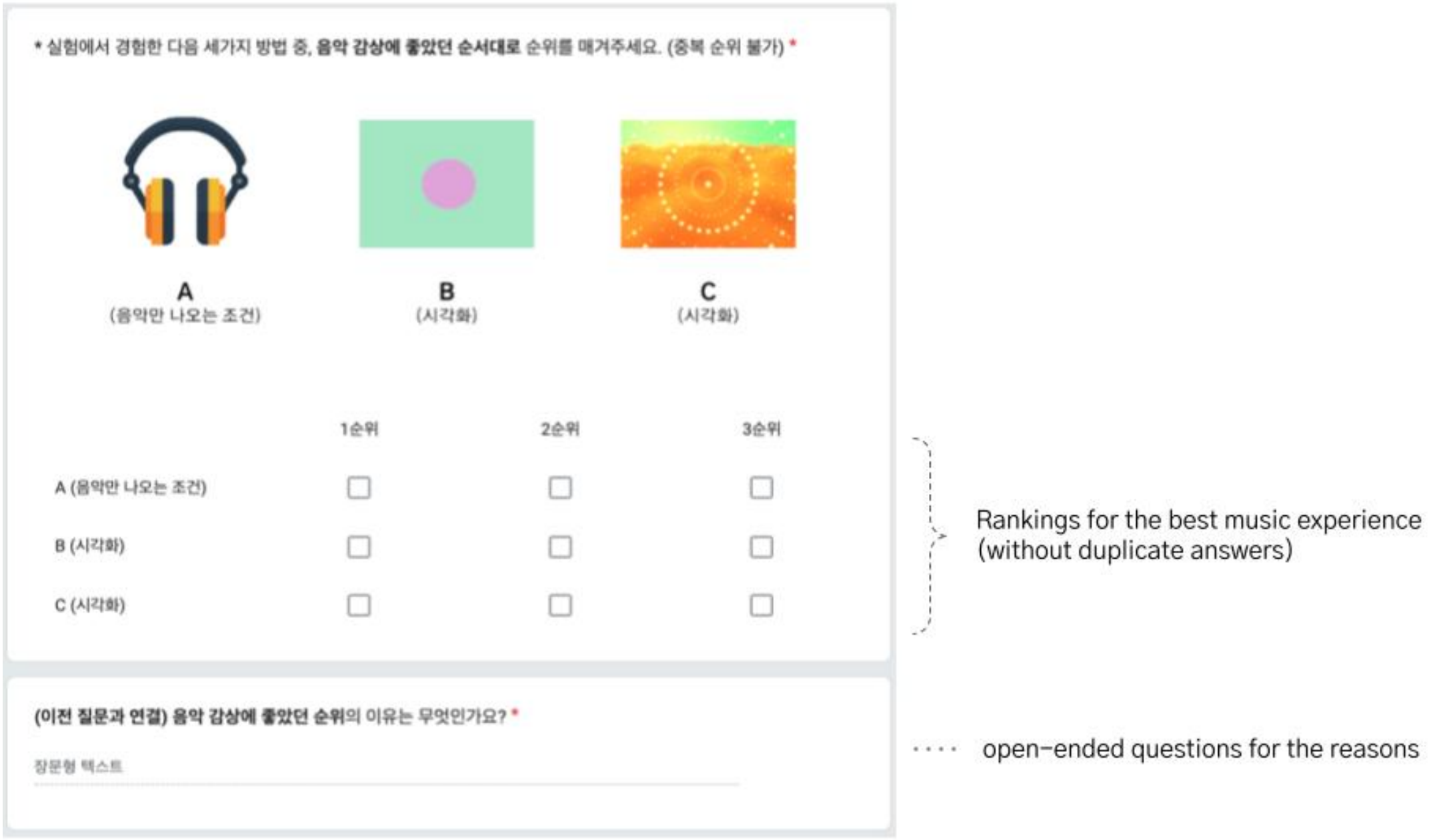

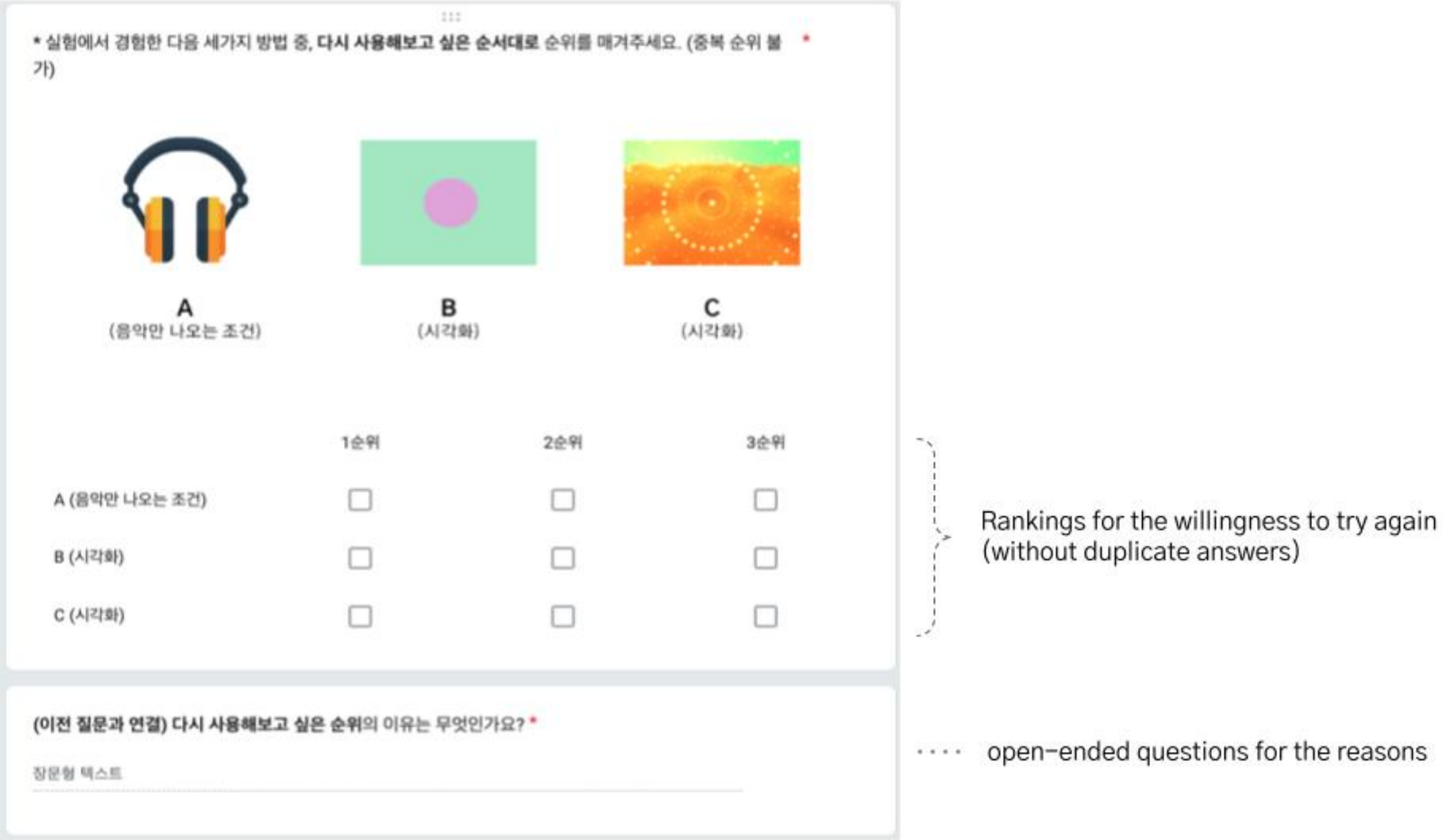

* 실험에서 경험한 다음 세가지 방법 중, **다시 사용해보고 싶은 순서대로** 순위를 매겨주세요. (중복 순위 불가) *

A (음악만 나오는 조건)

B (시각화)

C (시각화)

| | 1순위 | 2순위 | 3순위 |
|---|---|---|---|
| A (음악만 나오는 조건) | ☐ | ☐ | ☐ |
| B (시각화) | ☐ | ☐ | ☐ |
| C (시각화) | ☐ | ☐ | ☐ |

**(이전 질문과 연결) 다시 사용해보고 싶은 순위**의 이유는 무엇인가요? *

장문형 텍스트

# Curriculum Vitae

## ChungHa Lee M.S. Student

GIST IIT School of Integrated Technology,

Soft Computing & Interaction Laboratory

: chunghalee@gist.ac.kr

: https://github.com/ChungHaLee

Contact: +82-10-2667-6489

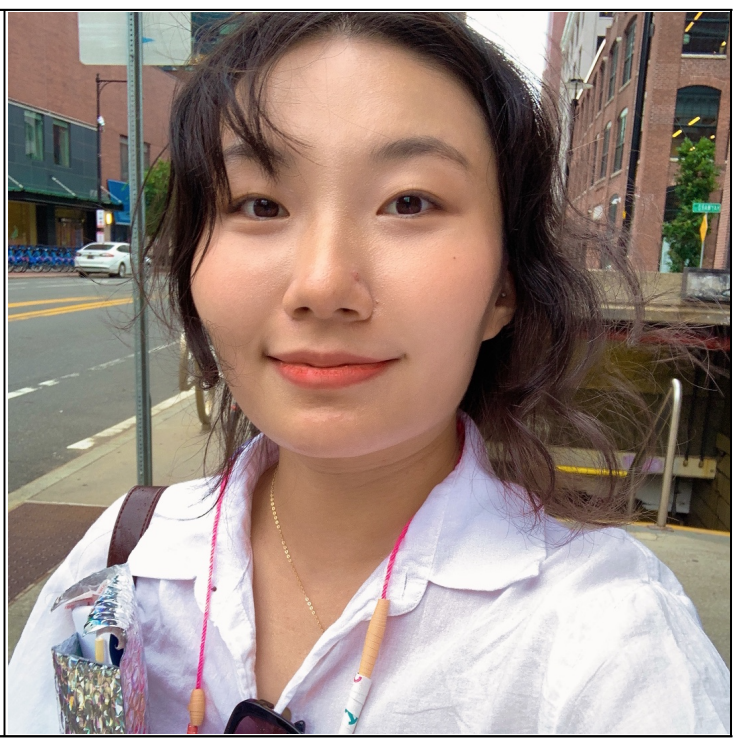

## Education

**M.S.,** Culture Technology (HCI), 2023, Gwangju Institute of Science and Technology.

**B.Ed.,** Pedagogy, 2021, Korea National University of Education.

History Education (Double Major), 2021, Korea National University of Education.

## Skills

Javascript, Python, User Experience and Interface

a good writer, a communicator for problem-solving, a fast learner

## Research Interests

HCI, Visualization, Application of Role-Playing Games in Education, User Centered Design

## Publications

1. **ChungHa Lee**, YouJin Choi, Junryeol Jeon, Jin-Hyuk Hong. **(2022).** A Study on the Music Visualization Tool for Deaf and Hard-of-Hearing: From the Perspective of Exploration and Customization, Proceedings of the Korean Information Science Society Conference. 1474-1476.
2. YouJin Choi, Junryeol Jeon, **ChungHa Lee**, Yeo-Gyeong Noh, Jin-Hyuk Hong. **(2022).** Cross-modal Music Palette: A Music Conceptualization Tool for the Deaf and Hard of Hearing to Enjoy Music. Under Review.
3. YouJin Choi, **ChungHa Lee**, Jin-Hyuk Hong. **(2021).** Design and Development of an Emotion Annotation System for Deaf and Hard-of-Hearing. Proceedings of the Korean Information Science Society Conference. 974-976.
4. JooYeong Kim, **ChungHa Lee**, JuYeon Kim, Jin-Hyuk Hong. **(2021).** Interactive Description to Enhance Accessibility and Experience of Deaf and Hard-of-Hearing Individuals in Museums. Under Review.

## Projects

1. Development of Assistive Technology of Music and Dance for Deaf and Hard-of-Hearings to Entertain Music. **(2022. 03 - Present)**
2. HCI+AI Convergence for Human-Centered Physical System Design. **(2022. 03 - Present)**

3. Development of Intelligent Exhibition Labels with Korean Sign Language Conversion Technology for Deaf and Hard-of-Hearings. (2021. 03 - 2022. 03)

## Awards and Certificate

2018. 02 NAU (Northern Arizona University) Education Program
2020. 01 DCU (Dublin City University) English Language Training
2020. 08 Grand Prize, POSCO Artificial Intelligence & Big Data Academy

## Extracurriculars

**KNUE Education Donors,** ***Team Leader of Multicultural Education***
2016. 07. Education Donation Acitivies at Jinbo High School, Republic of Korea.
2016. 08. Multicultural Eoulim Camp at Daeso Elementary School, Republic of Korea.
2016. 12. Education Donation Acitivies at Jinhae Girls' High School, Republic of Korea.
2017. 08. Multicultural Eoulim Camp at Daeso Elementary School, Republic of Korea.
2018. 01. Education Donation Acitivies at Gaeryung Middle School, Republic of Korea.
2018. 01. Multicultural Eoulim Camp at Deoksan Elementry School, Republic of Korea.